\documentclass[12pt]{article} % Default font size is 12pt, it can be changed here

\newcommand{\blind}{0}

\usepackage[
top    = 1.2in,
bottom = 1.4in,
left   = 1.0in,
right  = 1.0in]{geometry}
\usepackage{setspace}
\usepackage{times}
\usepackage{graphicx,psfrag,epsf}
\usepackage{amsmath}
\usepackage{amssymb}
\usepackage{gensymb}
\usepackage{subcaption}
\usepackage[toc,page]{appendix}
\usepackage{natbib}
\usepackage{xcolor}

\usepackage{graphicx} % Required for including pictures

\usepackage{url}
\usepackage{booktabs}
\usepackage{array}
\usepackage{multirow}

\usepackage{enumitem}

\usepackage{verbatim}
\usepackage[format=plain, labelfont=bf, singlelinecheck=off]{caption} 

\usepackage{algorithm}% http://ctan.org/pkg/algorithms
\usepackage{algpseudocode}% http://ctan.org/pkg/algorithmicx
\newlength{\continueindent}
\usepackage{etoolbox}
\makeatletter
\newcommand*{\ALG@customparshape}{\parshape 2 \leftmargin \linewidth \dimexpr\ALG@tlm+\continueindent\relax \dimexpr\linewidth+\leftmargin-\ALG@tlm-\continueindent\relax}
\apptocmd{\ALG@beginblock}{\ALG@customparshape}{}{\errmessage{failed to patch}}
\makeatother
\numberwithin{equation}{section}

\newcommand{\bfc}{\mathbf{c}}

\newcommand{\bfs}{\mathbf{s}}
\newcommand{\bfu}{\mathbf{u}}

\newcommand{\bfB}{\mathbf{B}}
\newcommand{\bfC}{\mathbf{C}}
\newcommand{\bfD}{\mathbf{D}}
\newcommand{\bfE}{\mathbf{E}}

\newcommand{\bfG}{\mathbf{G}}
\newcommand{\bfH}{\mathbf{H}}

\newcommand{\bfJ}{\mathbf{J}}
\newcommand{\bfK}{\mathbf{K}}
\newcommand{\bfL}{\mathbf{L}}
\newcommand{\bfM}{\mathbf{M}}

\newcommand{\bfP}{\mathbf{P}}
\newcommand{\bfQ}{\mathbf{Q}}
\newcommand{\bfR}{\mathbf{R}}
\newcommand{\bfS}{\mathbf{S}}
\newcommand{\bfT}{\mathbf{T}}
\newcommand{\bfU}{\mathbf{U}}
\newcommand{\bfV}{\mathbf{V}}
\newcommand{\bfW}{\mathbf{W}}
\newcommand{\bfX}{\mathbf{X}}
\newcommand{\bfY}{\mathbf{Y}}
\newcommand{\bfZ}{\mathbf{Z}}

\newcommand{\bfalpha}{\boldsymbol \alpha}

\newcommand{\bfbeta}{\boldsymbol \beta}
\newcommand{\bfepsilon}{\boldsymbol \epsilon}
\newcommand{\bfdelta}{\boldsymbol \delta}

\newcommand{\bfeta}{\boldsymbol \eta}
\newcommand{\bfxi}{\boldsymbol \xi}
\newcommand{\bftheta}{\boldsymbol \theta}

\newcommand{\bfDelta}{\boldsymbol \Delta}

\newcommand{\bfSigma}{\boldsymbol \Sigma}

\def\var{\mbox{\textrm{var}}}

\newcommand{\diag}{\text{diag}}

\expandafter\def\expandafter\normalsize\expandafter{%
    \normalsize
    \setlength\abovedisplayskip{6pt}
    \setlength\belowdisplayskip{6pt}
    \setlength\abovedisplayshortskip{6pt}
    \setlength\belowdisplayshortskip{6pt}
}

\usepackage{mathtools}

\begin{document}
\date{}

\def\spacingset#1{\renewcommand{\baselinestretch}%
{#1}\small\normalsize} \spacingset{1}

%%%%%%%%%%%%%%%%%%%%%%%%%%%%%%%%%%%%%%%%%%%%%%%%%%%%%%%%%%%%%%%%%%%%%%%%%%%%%%

\if0\blind
{
  \title{\bf Spatio-Temporal Data Fusion for Massive Sea Surface Temperature Data from MODIS and AMSR-E Instruments
  }

  \author{Pulong Ma\thanks{\emph{Correspondence to}: Pulong Ma,  The Statistical and Applied Mathematical Sciences Institute, 79 T.W. Alexander Drive, P.O. Box 110207, Durham, NC 27709. Email:  pma@samsi.info} \\
 Statistical and Applied Mathematical Sciences Institute  \\
  and Duke University \\
  %79 T.W. Alexander Drive, P.O. Box 110207 \\
  %Durham, NC 27709 \\
 % \textit{pma@samsi.info} \\
    Emily L. Kang \\
    University of Cincinnati\\
    %Cincinnati, OH 45220 \\
    %\textit{kangel@ucmail.uc.edu} \\
    }    

% (\textit{pma@samsi.info}; \textit{kangel@ucmail.uc.edu}) &    (\textit{\{amy.braverman, hai.nguyen\}@jpl.nasa.gov}) 

  \maketitle

} \fi

\if1\blind
{
  \bigskip
  \bigskip
  \bigskip
  \begin{center}
    {\LARGE\bf Spatio-Temporal Data Fusion for Massive Sea Surface Temperature Data from MODIS and AMSR-E Instruments}
\end{center}
  \medskip
} \fi

\bigskip
\begin{abstract}
Remote sensing data have been widely used to study various geophysical processes. With the advances in remote-sensing technology, massive amount of remote sensing data are collected in space over time. Different satellite instruments typically have different footprints, measurement-error characteristics, and data coverages. To combine datasets from different satellite instruments, we propose a dynamic fused Gaussian process (DFGP) model that enables fast statistical inference such as filtering and smoothing for massive spatio-temporal datasets in a data-fusion context. Based upon a spatio-temporal-random-effects model, the DFGP methodology represents the underlying true process with two components: a linear combination of a small number of basis functions and random coefficients with a general covariance matrix, together with a linear combination of a large number of basis functions and Markov random coefficients. To model the underlying geophysical process at different spatial resolutions, we rely on the change-of-support property, which also allows efficient computations in the DFGP model. To estimate model parameters, we devise a computationally efficient stochastic expectation-maximization (SEM) algorithm to ensure its scalability for massive datasets. The DFGP model is applied to a total of 3.7 million sea surface temperature datasets in the tropical Pacific Ocean for a one-week time period in 2010 from MODIS and AMSR-E instruments. 
\end{abstract}

\noindent%
{\it Keywords:} Dynamic fused Gaussian process; Spatio-temporal data fusion; Basis functions; Change of support; Massive datasets; Sea surface temperature
\vfill
%\hfill {\tiny technometrics tex template (do not remove)}

\newpage
%\spacingset{1.6} % DON'T change the spacing!
\doublespacing
%----------------------------------------------------------------------------------------
%	Introduction
%----------------------------------------------------------------------------------------

\section{Introduction}
Remote sensing technology has been advancing the measurement of massive amount of datasets for many geophysical processes. Statistical analysis for massive amount of data is challenging, since many geophysical processes evolve in space and time with complicated structures. The resulting data often exhibit nonstationary dependence structures. As remote sensing data are often collected by different satellite instruments over different footprints that have distinct shapes, orientations and sizes, these remote-sensing data are often noisy and incomplete with incompatible spatial supports and distinct measurement-error characteristics. In our real application, we focus on data collected from polar-orbiting satellites. A key characteristic of remote sensing data from polar-orbiting satellites is that these data are often collected at very high spatial resolution and relatively low temporal resolution. Spatial modeling for such data even at a single time point is computationally challenging, which leads to the well-known big ``$n$'' problem in spatial statistics \citep{Cressie2006, Cressie2008, Banerjee2008, Nychka2015, Ma2017}. In a spatio-temporal setting with more than one data sources, it is much more challenging to tackle this computational issue. 

To analyze data from different satellite instruments, the resolution difference must be accounted for. There is a vast literature in spatial statistics to tackle the so-called change-of-support problem when statistical analysis is carried out with several data sources at different resolutions. Here, change of support (COS) refers to inference made at a resolution based on data from different scales \citep[e.g.,][]{Cressie1993, Cressie1996, Gelfand2001, Gotway2002}. A direct way to deal with the change-of-support problem is to represent the process at the block level as a stochastic integral of the process at the point level or areal-unit level. When data are obtained at different scales or resolutions, statistical inference for combining such data leads to the data-fusion problem. In spatial-statistics literature, data fusion has been approached in several different ways. \cite{Wikle2005} formulate a hierarchical Bayesian model that combines observations across different scales by assuming the same underlying true process. In a similar way, \cite{Fuentes2005} present an instance of Bayesian melding \citep{Poole2000} that assumes the underlying true process at the point level so that point-referenced observations of air pollution data and air-quality model output at the grid-cell level are linked to a same true process at the point level via measurement-error processes. The space-time extension of \cite{Fuentes2005} has been developed in \cite{Choi2009} to study the spatio-temporal association between mortality and pollution exposure to daily fine particulate matter (PM$_{2.5}$) based on  point-referenced PM$_{2.5}$ and air quality model output. These models address the change-of-support problem explicitly, but their implementations require expensive computations, and hence their models are not suitable to directly analyze large datasets. Other different approaches for data fusion have been developed as well. For instance, \cite{McMillan2010} present a spatio-temporal model that combines point-referenced observations and numerical model output at the grid-cell level by assuming that the data process at point level is linked to the same underlying true process at the grid-cell level. Instead of assuming the same true underlying process for both observations and numerical model output, \cite{Berrocal2010, Berrocal2012datafusion} propose to regress the point-referenced observations over model output at the grid-cell level with spatial or spatio-temporal varying coefficients. These models have been applied to make inference based on point-referenced air quality observations and numerical model output at grid-cell level over the United States. In this paradigm, \cite{Sahu2010} also regress the point-referenced true process over numerical model output at the grid-cell level to predict chemical deposition in the eastern United States. These approaches do not address the change-of-support explicitly, and they require intensive computations to fit the model in a Bayesian framework. In remote sensing science, massive amount of data are often collected over space and time by satellite instruments, making these methods computationally intractable. 
 
To tackle the massiveness of remote-sensing data, \cite{Nguyen2012} present the spatial data fusion methodology based on the spatial-random-effects model \citep{Cressie2006, Cressie2008}, where a single underlying true spatial process is assumed at the areal-unit level. \cite{Nguyen2014} further develop the spatio-temporal data fusion methodology based on the spatio-temporal-random-effects model \citep{Cressie2010, Kang2010, Katzfuss2011, Katzfuss2012}, where different underlying true spatio-temporal processes are assumed at the areal-unit level, and cross-dependence structures among different true processes are modeled through the spatio-temporal-random-effects model. 

In this article, we propose a dynamic fused Gaussian process (DFGP) methodology for spatio-temporal data fusion to combine multiple datasets from different satellite instruments. As a generalization of \cite{Nguyen2014}, our DFGP methodology extends the spatial-only fused Gaussian process (FGP) in \cite{Ma2017} to a spatio-temporal setting. In particular, the FGP model extends the fixed rank kriging model \citep{Cressie2006, Cressie2008} by combining a low-rank representation with a general covariance matrix together with a graphical model with a sparse precision matrix. Based upon FGP, we take a dynamic-statistical approach to build the DFGP model under which the current state of the process of interest evolves from the previous state in a dynamic way. This hierarchical modeling approach has been adopted in many previous research \citep[e.g.,][]{Mardia1998, Wikle1998, Wikle1999, Berliner2000, Stroud2001, Wikle2001, Huang2002, Cressie2002, Xu2007}; see \cite{Cressie2011} for a comprehensive overview for spatio-temporal models. Our proposed DFGP falls into this paradigm, and extends the spatio-temporal-random-effects model \citep{Cressie2010, Kang2010, Katzfuss2011, Katzfuss2012, Zammit2017} with a more flexible covariance function.

The reminder of this article is organized as follows. Section~\ref{sec: data} introduces two different datasets from two satellite instruments onboard NASA's AQUA satellite. Section~\ref{sec: DFGP} presents the dynamic fused Gaussian process methodology in a data-fusion context. Kalman filtering and Kalman smoothing procedures are also derived. In Section~\ref{sec: SEM}, we give details on the stochastic expectation-maximization algorithm for parameter estimation in both filtering and smoothing procedures. In Section~\ref{sec: results}, we apply the DFGP methodology to analyze massive amount of sea surface temperature datasets, and make comparisons with other existing methods such as the spatio-temporal data fusion model in \cite{Nguyen2014}. Section~\ref{sec: conclusion} concludes with discussions and future research work.

\section{Data} \label{sec: data}
Sea surface temperature (SST) is a key climate and weather measurement, which plays a crucial role in understanding climate systems. Massive amount of SST datasets are collected from satellite instruments each day with the advances in new remote-sensing technology. For instance, the AQUA satellite launched on May 4, 2002 is a polar-orbiting satellite around the Earth, aiming at studying Earth's precipitation, evaporation, and cycling of water. The AQUA satellite carries two instruments: the Moderate Resolution Imaging Spectroradiometer (MODIS) and the Advanced Microwave Scanning Radiometer-Earth Observing System (AMSR-E). The MODIS instrument is an infrared radiometer with a ground swath width of 2,330 km, which is able to measure SST at fine spatial resolutions, but is unable to measure through cloud cover; the AMSR-E instrument is a microwave radiometer with a ground swath width of 1445 km, which is able to measure SST in all weather conditions except rain, but only at coarse spatial resolutions, though its quality is also subject to radio frequency interference. MODIS SST and AMSR-E SST data have been widely used for scientific research and operations \citep[e.g.,][]{Donlon2002, Carroll2006, Gentemann2014}, However, it is still challenging to combine these two different data products due to their different characteristics including different spatial resolutions. \textit{Heuristic} methods and empirical comparisons with in-situ observations are studied \citep[e.g.,][]{Guan2003, Kawai2006, Arai2013}. In this article, we propose a spatio-temporal statistical model to generate such high-resolution SST data products on a daily scale by combining MODIS SST and AMSR-E SST data in a rigorous way. The resulting high-resolution SST products can be critical and helpful for operational oceanography and numerical weather prediction.

\begin{figure}[htbp]
\begin{center}
\makebox[\textwidth][c]{ \includegraphics[width=1.0\textwidth, height=0.4\textheight]{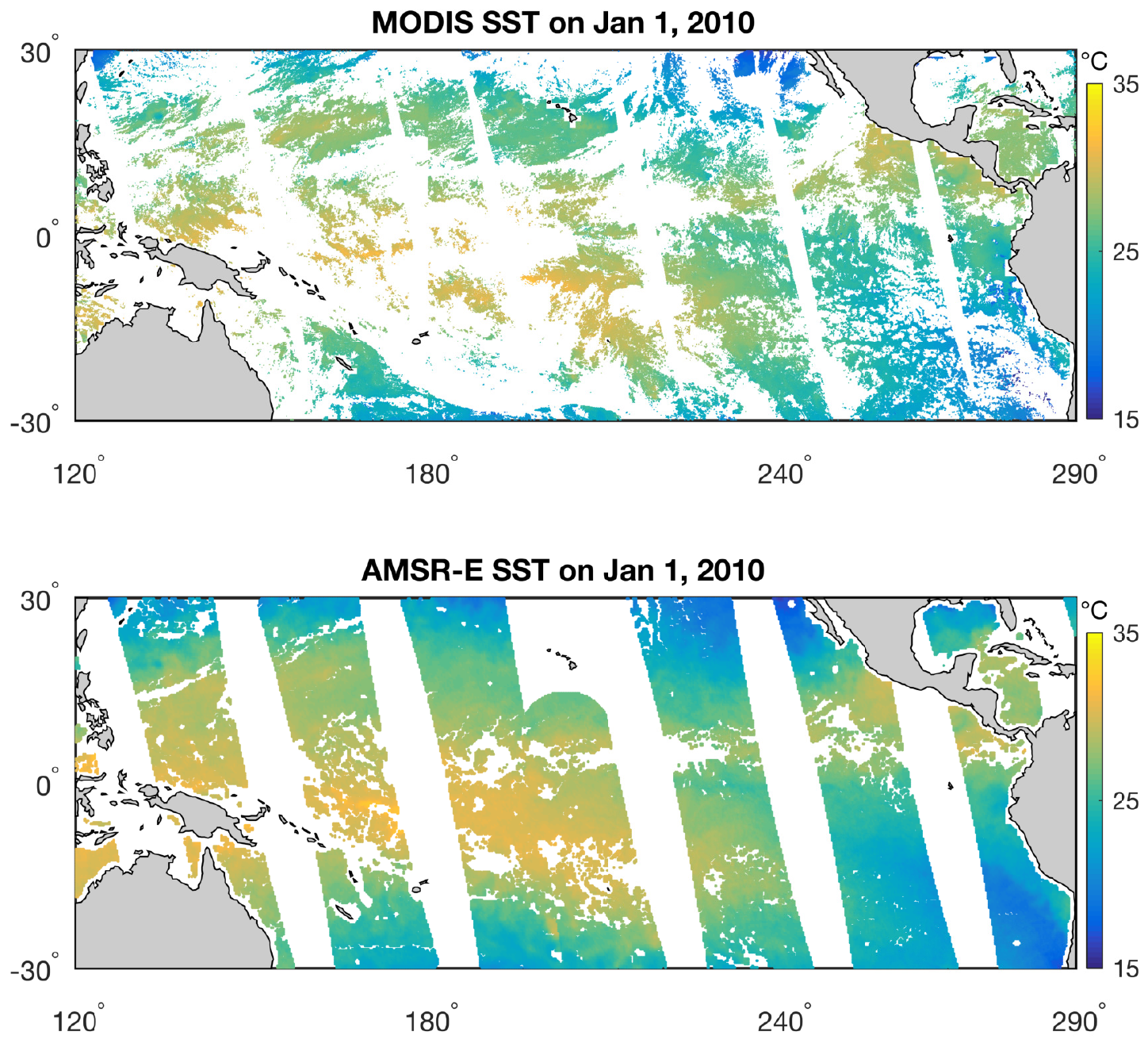}}
\caption{MODIS and AMSR-E SST data on January 1, 2010. The MODIS SST data are at 9 km resolution and the AMSR-E SST data are at 25 km resolution.}
\label{fig: MODIS_AMSRE_day1}
\end{center}
\end{figure}

In this study, we use daily daytime MODIS SST data at 9 km spatial resolution processed from the NASA Ocean Biology Processing Group Data Center (\url{oceancolor.gsfc.nasa.gov}), and daily daytime AMSR-E SST data at 25 km spatial resolution from \url{www.remss.com} from January 1 to 8 in the year 2010. These datasets have distinct error characteristics and are often sparse, irregular, and noisy with incompatible supports. Statistical methods for combining different sources of remote-sensing data will give much more accurate and reliable uncertainty analysis. The study region is chosen to be the tropical Pacific region  between longitude $-30^{\circ}$ and $30^{\circ}$ and between latitude $120^{\circ}$ and $290^{\circ}$ from January 1 to 8 in the year 2010. Figure~\ref{fig: MODIS_AMSRE_day1} shows the MODIS SST and AMSR-E SST on January 1, 2010. The numbers of observations from MODIS instrument for each day are $n_1^{(1)} = 362,721, n_2^{(1)} = 398,662, n_3^{(1)}=409,445, n_4^{(1)}=385,490, n_5^{(1)}=425,541, n_6^{(1)}=415,869, n_7^{(1)}=416,721, n_8^{(1)}=415,467$, respectively, resulting in a total of 3,229,916 MODIS SST observations, where the subscript denotes the time step, and the superscript denotes the instrument with 1 for MODIS and 2 for AMSR-E. The numbers of observations from AMSR-E instrument for each day are $n_1^{(2)}=71,592, n_2^{(2)}=68,574, n_3^{(2)}=72,905, n_4^{(2)}=64,868, n_5^{(2)}=73,228, n_6^{(2)}=66778, n_7^{(2)}=71,431, n_8^{(2)}=67,245$, respectively, resulting in a total of 556,621 AMSR-E SST observations. 

\section{The Dynamic Fused Gaussian Process Model} \label{sec: DFGP}
For many physical spatio-temporal processes, it is quite natural that the process of interest cannot be observed directly, and we assume that the data process is a sum of a hidden process and a measurement-error process. Suppose we are interested in a real-valued spatio-temporal process $\{Y_t(\bfs): \bfs \in \mathcal{D}\subset \mathbb{R}^d,\, t\in \mathcal{T}\}$, where $\mathcal{D} \subset \mathbb{R}^d$ and $\mathcal{T}\equiv \{1, 2, \ldots, T\}$ for a positive integer $T$. The goal is to make fast statistical inference such as filtering and smoothing based on massive spatio-temporal datasets. In what follows, we present the dynamic fused Gaussian process (DFGP) model. 

Suppose that the spatial domain $\mathcal{D}$ of interest is made up of $N$ non-overlapping, equal-areal, basic areal units $\mathcal{R}_i$ for $i=1, \ldots, N$, where each $\mathcal{R}_i$ is assumed to be associated with its centroid $\bfs_i$. These basic areal units represent the smallest spatial resolution at which prediction will be made, and they are called BAUs. In what follows, the discretized version of the domain $\mathcal{D}$ will be referred to as $\mathcal{R} \equiv \cup_{i=1}^N\{\mathcal{R}_i\}$ that is indexed by corresponding centroids $\{ \bfs_i: i=1, \ldots, N\}$. This discretization procedure has been used in many previous work \citep[e.g.,][]{Cressie2006, Cressie2008, Nguyen2012, Nguyen2014, Shi2017, Ma2017, Ma2018downscaling}. The data process is assumed to be observed at different spatial resolutions resulted from the same underlying true process $Y_t(\cdot)$ over these BAUs. This is a typical situation where multiple satellite instruments measure the same geophysical process at different resolutions in remote sensing science. Let $\bfZ_t^{(k)} \equiv (Z_t(\mathcal{A}_{t,1}^{(k)}), \dots, Z_t(\mathcal{A}_{t, n_t^{(k)}}^{(k)}))'$ be a vector of noisy version of the underlying true process $Y_t(\cdot)$ at the $k$-th spatial resolution over $n_t^{(k)}$ footprints $\{ \mathcal{A}_{t,i}^{(k)}: i=1, \ldots, n_t^{(k)}\}$ collected from the $k$-th satellite instrument at time $t$, where the quantity with superscript $k$ corresponds to that from the $k$-th satellite instrument. The total number of observations at time $t$ is denoted as $n_t = \sum_{k=1}^{k_0} n_t^{(k)}$ across all resolutions with $k_0$ being the number of instruments. The data process $Z_t(\cdot)$ over the $i$-th footprint $\mathcal{A}_{t,i}^{(k)}$ from $k$-th instrument at time $t$ is modeled as the true process $Y_t(\cdot)$ over footprint $\mathcal{A}_{t,i}^{(k)}$ plus a measurement-error process:

\begin{eqnarray}\label{eqn: data model in DFGP}
Z_t(\mathcal{A}_{t,i}^{(k)}) = Y_t(\mathcal{A}_{t,i}^{(k)}) + \epsilon_t(\mathcal{A}_{ t,i}^{(k)}), \quad \mathcal{A}_{ t,i}^{(k)} \subset \mathcal{D}; i=1, \ldots, n_t^{(k)}; k=1, \ldots, k_0
\end{eqnarray}
The true process $Y_t(\cdot)$ over the footprint $\mathcal{A}_{ t,i}^{(k)}$ is assumed to be the block average of the process $Y_t(\cdot)$ over the BAUs within the domain $\mathcal{A}_{ t,i}^{(k)}$:
%\begin{eqnarray} \label{eqn: cosp in DFGP}
%Y_t (\mathcal{A}) = \frac{1}{|\mathcal{A}|} \int_{\mathbf{s} \in \mathcal{A}} Y(\bfs)\, d\bfs. 
%Y_t (\mathcal{A}_{t,i}^{(k)}) = \frac{1}{|\mathcal{A}_{t,i}^{(k)} \cap \mathcal{D}|} \int_{\mathcal{A}_{t,i}^{(k)}} Y(\bfs)\, d\bfs. 
%\end{eqnarray}
\begin{eqnarray} \label{eqn: discrete sum in DFGP}
Y_t (\mathcal{A}_{t,i}^{(k)}) = \frac{1}{\sum_{\ell=1}^N I(\bfs_{\ell} \in \mathcal{A}_{t,i}^{(k)})} \sum_{\ell=1}^N I(\bfs_{\ell} \in \mathcal{A}_{t,i}^{(k)})\cdot Y(\bfs_{\ell}).
\end{eqnarray} 
where $\bfs_{\ell}$ is the centroid of the BAU $\mathcal{R}_{\ell}$. The Eq.~\eqref{eqn: discrete sum in DFGP} is accounting for the change-of-support problem with BAUs assumed to have equal areas. This strategy has been widely taken in previous work \citep[e.g.,][]{Nguyen2012, Nguyen2014, Nguyen2017, Ma2018downscaling}.

The measurement-error process $\epsilon_t(\cdot)$ is assumed to be a Gaussian white noise term and $\{\epsilon_t(\mathcal{A}_{ t,i}^{(k)});$ $i=1, \ldots, n_t^{(k)}; k=1, \ldots, k_0; t=1, \ldots, T\}$ are assumed to be independent. It may have nonzero mean capturing the instrument bias, and has variance $\text{var}(\epsilon_t(\mathcal{A}_{ t,i}^{(k)}))= \sigma_{\epsilon_t, (k)}^{2} v(\mathcal{A}_{ t,i}^{(k)})>0$, where $v(\mathcal{A}_{ t,i}^{(k)})$ is known from validation data and instrument specification and allows for the possibility of nonconstant variance. The variance parameter $\sigma_{\epsilon_t, (k)}^{2}$ will be estimated via a maximum likelihood estimation procedure in Section~\ref{sec: SEM}. Under this assumption, the nugget variance parameters are both instrument-dependent and time-dependent. This allows great flexibility in modeling the measurement-error processes. The data model defined by Eq.~\eqref{eqn: data model in DFGP} and \eqref{eqn: discrete sum in DFGP} has been studied in many previous work \citep[e.g.,][]{Nguyen2012, Nguyen2014}. 

As we are interested in the process $Y_t(\cdot)$ at the finest resolution defined by $N$ BAUs. Following \cite{Nguyen2012, Nguyen2014, Ma2018downscaling}, we define the process of interest $\{Y_t(\bfs): \bfs \in \mathcal{R}\}$ at BAU-level:
\begin{eqnarray} 
 Y_t(\bfs) &=& \bfX_t(\bfs)' \bfbeta_t + \nu_t(\bfs) + \delta_t(\bfs), \label{eqn: model for Y}\\
 \nu_t(\bfs) &=& \bfS_t(\bfs)'\bfeta_t, \label{eqn: model for nu} \\
 \delta_t(\bfs) &=& \bfB_t(\bfs)' \bfxi_t, \label{eqn: model for delta}
\end{eqnarray}
where $\bfX_t(\cdot)\equiv (X_{t,1}(\cdot), \ldots, X_{t,p}(\cdot))'$ is a $p$-dimensional vector of covariates and $\bfbeta_t$ are corresponding unknown coefficients at time $t$; $\bfS_t(\cdot) \equiv (S_{t,1}(\cdot), \ldots, S_{t,r}(\cdot))'$ is an $r$-dimensional vector of basis functions at time $t$, and $\bfeta_t$ is an $r$-dimensional random vector. The quantities $\bfX_t(\cdot), \bfS_t(\cdot),  \bfB_t(\cdot)$ are defined at point level, and then are computed at BAU-level using Monte Carlo techniques following previous work \citep{Wikle2001, Gelfand2001, Fuentes2005, Katzfuss2011, Katzfuss2012}. More specifically, these quantifies at BAU-level can be approximated by Monte Carlo averages of them defined over 30 uniformly-distributed point locations within each BAU. In what follows, the pre-specified quantities defined at BAU-level are always obtained from their specification at point level via such Monte Carlo techniques.

Following \cite{Cressie2008}, we choose a bisquare basis function with the following form: $S_{t,i}(\bfu) = \{1 - (\|\bfu - \bfc_i\|/\ell_i)^2 \}^2\cdot\{I(\|\bfu - \bfc_i\| \leq \ell_i)\}$ for $i=1, \ldots, r$, where $\bfc_i$ is the center of the $i$-th bisquare basis function, and $\ell_i$ is the corresponding radius of the $i$-th bisquare basis function. In addition, we also assume that the number of basis functions is much smaller than the number of observations, i.e., $r\ll n_t$. The model in Eq~\eqref{eqn: model for nu} has a low-rank representation. $\bfB_t(\cdot)\equiv (B_{t,1}(\cdot), \ldots, B_{t, N}(\cdot))'$ is an $N$-dimensional vector of basis functions for the Markov random coefficients $\bfxi_t$ at time $t$. $\bfxi_t$ is an $N$-dimensional random vector defined on $N$ BAUs with Markov structure specified in Eq.~\eqref{eqn: model for xi}. Following \cite{Ma2017}, we choose a piecewise constant basis function for $B_{t,i}(\cdot)$ with the following form: $\bfB_{t,i}(\bfu)=I(\bfu \in \mathcal{R}_i)$ for $i=1, \ldots, N$. The model in Eq.~\eqref{eqn: model for delta} has a high-rank representation, since $N\approx n_t$ or $N>n_t$. Notice that the quantities $\bfX_t(\cdot), \bfS_t(\cdot), \bfB_t(\cdot)$ defined at BAU-level lead to a model for the underlying true process $Y_t(\cdot)$ also defined at BAU-level. However, the data are collected at a resolution that is coarser than the resolution at which these BAUs are defined. The Eq.~\eqref{eqn: discrete sum in DFGP} links the process $Y_t(\cdot)$ at BAU-level to the resolution at which the data process is defined through the change-of-support property. In what follows, we give the model specification for $\bfeta_t$ and $\bfxi_t$.
%Pre-multiplying them by the weight vector $\bfA(\cdot)$ will give corresponding quantities defined at coarse spatial resolutions. %For notational convenience, the following conventions will be used: $\bfX_t(\mathcal{A}) \equiv \bfPhi(\mathcal{A})' \bfX_t(\bfs)$; $\bfS_t(\mathcal{A}) \equiv \bfPhi(\mathcal{A})' \bfS_t(\bfs)$; $\bfA_t(\mathcal{A}) \equiv \bfPhi(\mathcal{A})' \bfA_t(\bfs)$. Let $\bfX_t^{(k)}\equiv [\bfX_t(\mathcal{A}_1^{(k)}),\ldots, \bfX_t(\mathcal{A}_{n_t^{(k)}}^{(k)})]'$ denote a vector of covariates at all $n_t^{(k)}$ footprints from $k$-th instrument at time point $t$.

Following \cite{Cressie2010},  we assume that the dynamical evolution of $\{\bfeta_t: t=0, 1, \ldots, T\}$ follows a vector-autoregressive (VAR) model of order 1:
\begin{eqnarray} \label{eqn: evolution model}
\bfeta_t \mid \bfeta_{t-1}, \bfeta_{t-2}, \ldots, \bfeta_0 \sim \mathcal{N}_r(\mathbf{H}_t \bfeta_{t-1}, \mathbf{U}_t),\, t=1, 2, \ldots, T,
\end{eqnarray}
with the initial state $\bfeta_0 \sim \mathcal{N}_r(\mathbf{0},\, \bfK_0)$. The $r\times r$ matrix $\bfH_t$ and $r\times r$ matrix $\bfU_t$ are referred to as the propagation matrix and innovation covariance matrix, respectively.

The spatial-temporal process $\delta_t(\cdot)$ is linked to the random vector $\bfxi_t$ through the link matrix $\bfB_t(\cdot)$. As the domain is partitioned into pairwise disjoint subregions $\{\mathcal{R}_i: i=1, \ldots, N\}$, we assume the following parsimonious spatial-temporal model for $\bfxi_t=(\xi_t(\mathcal{R}_1), \ldots, \xi_t(\mathcal{R}_N))'$: for $i=1, \ldots, N$, 
\begin{eqnarray} \label{eqn: model for xi}
\xi_t(\mathcal{R}_i) \mid \bfxi_t^{-i} \sim \mathcal{N}(\gamma_{t} /e_{i+} \cdot \sum_{j\in \partial \mathcal{R}_i} e_{ij} \xi_t(\mathcal{R}_j),\, \tau_t^2/e_{i+}),
\end{eqnarray}
where $\bfxi_t^{-i}\equiv (\xi_t(\mathcal{R}_1), \ldots, \xi_t(\mathcal{R}_{i-1}), \xi_t(\mathcal{R}_{i+1}), \ldots, \xi_t(\mathcal{R}_{N}))'$; $\gamma_{t}$ is the spatial dependence parameter at time $t$; $\mathbf{E}=(e_{ij})$ is an $N \times N$ adjacency matrix on the discretized domain $\mathcal{R}$ at BAU-level, see Chapter 6 of \cite{Cressie1993} for various specifications; $\partial \mathcal{R}_i$ is a set of indices corresponding to the neighbors of the BAU $\mathcal{R}_i$ that are defined by the spatial adjacency matrix $\mathbf{E}$; and $e_{i+}=\sum_{j=1}^N e_{ij}$ for $i=1, \ldots, N$. $\tau^2_t$ is the conditional marginal variance parameter at time $t$.  From Eq.~\eqref{eqn: model for xi}, it is easy to derive that the joint conditional distribution of $\bfxi_t$ for $t=1, \ldots, T$, is
\begin{eqnarray} \label{eqn: conditional dist for xi}
\bfxi_t \sim \mathcal{N}(\mathbf{0},\, \bfQ_t^{-1}),
\end{eqnarray}
where $\bfQ_t \equiv \boldsymbol \bfDelta^{-1} (\mathbf{I} - \gamma_{t} \bfW)/\tau_t^2$; $\bfW \equiv\boldsymbol \bfDelta \cdot \bfE$ is the $N\times N$ proximity matrix, and $\boldsymbol \Delta \equiv \diag(1/ e_{1+}, \ldots, 1/e_{N+})$. This model is called a conditional autoregressive (CAR) model. The model for $\delta_t(\cdot)$ is a special case of the Gaussian graphical model (GGM). As discussed in \cite{Ma2017}, the model for $\delta_t(\cdot)$ can be constructed in a similar way as in \cite{Lindgren2011} and \cite{Nychka2015}. This will increase the flexibility of the model. Such implementation and demonstration is beyond the scope of this article. Although the model for $\delta_t(\cdot)$ does not incorporate dynamic evolution, the dynamic structure of the process $Y_t(\cdot)$ is inherited from the random vectors $\{\bfeta_t: t=1, \ldots, T\}$. 

Define the following quantities:
\begin{eqnarray}
X_{t, i}(\mathcal{A}_{t,j}^{(k)}) \equiv \frac{1}{\sum_{\ell=1}^N I(\bfs_{\ell} \in \mathcal{A}_{t,j}^{(k)})} \sum_{\ell=1}^N I(\bfs_{\ell} \in \mathcal{A}_{t,j}^{(k)})\cdot X_{t,i}(\bfs_{\ell}), \, i=1, \ldots, p, \\
S_{t, i}(\mathcal{A}_{t,j}^{(k)}) \equiv \frac{1}{\sum_{\ell=1}^N I(\bfs_{\ell} \in \mathcal{A}_{t,j}^{(k)})} \sum_{\ell=1}^N I(\bfs_{\ell} \in \mathcal{A}_{t,j}^{(k)})\cdot S_{t, i}(\bfs_{\ell}), \, i=1, \ldots, r, \\
B_{t,i}(\mathcal{A}_{t,j}^{(k)}) \equiv \frac{1}{\sum_{\ell=1}^N I(\bfs_{\ell} \in \mathcal{A}_{t,j}^{(k)})} \sum_{\ell=1}^N I(\bfs_{\ell} \in \mathcal{A}_{t,j}^{(k)})\cdot B_{t,i}(\bfs_{\ell}), \, i=1, \ldots, N,
\end{eqnarray}
where $t=1, \ldots, T$; $k=1, \ldots, k_0$; $j=1, \ldots n_t^{(k)}$. By combining Eq~\eqref{eqn: data model in DFGP} through Eq~\eqref{eqn: model for delta}, we have the following linear model at the $k$-th resolution:
\begin{eqnarray}
Z_t(\mathcal{A}_{t,i}^{(k)}) = \bfX_t(\mathcal{A}_{t,i}^{(k)})' \bfbeta_t + \bfS_t(\mathcal{A}_{t,i}^{(k)})' \bfeta_t + \bfB_t(\mathcal{A}_{t,i}^{(k)})' \bfxi_t + \bfepsilon_t(\mathcal{A}_{t,i}^{(k)}).
\end{eqnarray}
Now stacking the above model over all footprints from the $k$-th instrument at time $t$ yields the following representation: 
\begin{eqnarray}
\bfZ_t^{(k)} &=& \bfX_t^{(k)}\bfbeta_t + \bfS_t^{(k)} \bfeta_t + \bfB_t^{(k)} \bfxi_t + \bfepsilon_t^{(k)}, \quad k=1, \ldots, k_0,
\end{eqnarray}
where $\bfX_t^{(k)}\equiv[\bfX_t(\mathcal{A}_{t,1}^{(k)}), \ldots, \bfX_t(\mathcal{A}_{t,n_t^{(k)}}^{(k)})]'$ is an $n_t^{(k)}$-by-$p$ matrix.  $\bfS_t^{(k)}\equiv[\bfS_t(\mathcal{A}_{t,1}^{(k)}), \ldots, $  $\bfS_t(\mathcal{A}_{t,n_t^{(k)}}^{(k)})]'$ is an $n_t^{(k)}$-by-$r$ basis matrix in the low-rank component. $\bfB_t^{(k)}\equiv[\bfB_t(\mathcal{A}_{t,1}^{(k)}), \ldots, \bfB_t(\mathcal{A}_{t,n_t^{(k)}}^{(k)})]'$ is an $n_t^{(k)}$-by-$N$ basis matrix in the GGM component. $\bfepsilon_t^{(k)}\equiv [\epsilon_t(\mathcal{A}_{t,1}^{(k)}), \ldots,  \epsilon_t(\mathcal{A}_{t,n_t^{(k)}}^{(k)})]'$ is an $n_t^{(k)}$-dimensional random vector whose covariance matrix is $\sigma^2_{\epsilon_t, (k)} \bfV_{\epsilon_t, (k)}$ with  $\bfV_{\epsilon_t, (k)} \equiv\text{diag}\{v(\mathcal{A}_{t,1}^{(k)}), $ $ \ldots, v(\mathcal{A}_{t,n_t^{(k)}}^{(k)})\}$.

Let $\bfZ_t \equiv [\bfZ_t^{(1)'}, \ldots, \bfZ_t^{(k_0)'}]'$ be an $n_t$-dimensional vector stacking all the observations together from $k_0$ satellite instruments at time $t$. Let $\bfX_t \equiv [\bfX_t^{(1)'}, \ldots, \bfX_t^{(k_0)'}]'$ be an $n_t$-by-$p$ matrix stacking all the covariates corresponding to all the observations at time $t$. Let $\bfS_t \equiv [\bfS_t^{(1)'}, \ldots, \bfS_t^{(k_0)'}]'$ be an $n_t$-by-$r$ basis matrix in the low-rank component. Let $\bfB_t \equiv [\bfB_t^{(1)'}, \ldots, \bfB_t^{(k_0)'}]'$ be an $n_t$-by-$N$ basis matrix in the GGM component. Then we have 
\begin{eqnarray}
\bfZ_t = \bfX_t\bfbeta_t + \bfS_t \bfeta_t + \bfB_t \bfxi_t + \bfepsilon_t,
\end{eqnarray}
where $\bfepsilon_t\equiv (\bfepsilon_t^{(1)'}, \ldots, \bfepsilon_t^{(k_0)'})'$ is an $n_t$-dimensional random vector with covariance matrix $\bfV_t\equiv \text{diag}\{\sigma^2_{\epsilon_t, (1)} \bfV_{\epsilon_t, (1)},  \ldots, \sigma^2_{\epsilon_t, (k_0)} \bfV_{\epsilon_t, (k_0)}\}$.

%Without loss of generality, data from two different satellite instruments at two different footprint sizes are considered here, and hence $k_0=2$.

\subsection{Kalman Filter and Kalman Smoother} 
Suppose that inference is made on $Y_t(\bfs_0)$ for any centroid $\bfs_0$ of a BAU in $\mathcal{R}$ and any time $t=1, \ldots, T$. Let the set $\mathcal{R}^P$ be a collection of $m_t$ centroids where we want to make prediction of $Y_t(\cdot)$ at BAU-level for time $t$. So, the process vector of interest is 
\begin{eqnarray}
\bfY_t^P \equiv \bfX_t^P \bfbeta_t + \bfS_t^P \bfeta_t + \bfdelta_t^P, \, t=1, \ldots, T,
\end{eqnarray}
where superscript $P$ denotes the quantities evaluated at the set $\mathcal{R}^P$ of prediction locations. 

In what follows, sequential updates are given based on the hierarchical dynamical spatio-temporal process DFGP. The formulas are derived using Bayes' theorem in the context of dynamic spatio-temporal models described in \cite{Cressie2011}. To fix notation, we use $\bfZ_{1:u}\equiv [ \bfZ_1', \ldots, \bfZ_u']'$ to denote all the data collected from time $t=1$ to time $t=u$. For conditional expectations of $\bfeta_t$ and $\bfdelta_t^P$ based on $\bfZ_{1:u}$, the following notations will be used: $\bfeta_{t|u} \equiv E(\bfeta_{t}\mid \bfZ_{1:u})$ and $\bfdelta_{t|u}^P\equiv E(\bfdelta_t^P \mid \bfZ_{1:u})$. The corresponding conditional covariance matrices will be denoted as $\bfP_{t|u}\equiv \var(\bfeta_t\mid \bfZ_{1:u})$ and $\bfR_{t|u}^P\equiv \var(\bfdelta_t^P \mid \bfZ_{1:u})$, respectively.

Assuming initial states $\bfeta_{0|0} \equiv \mathbf{0}$ and $\bfP_{0|0} \equiv \bfK_0$, the one-step ahead forecast distribution $[\bfeta_t \mid \bfZ_{1:t-1}]$ is multivariate normal with mean  $\bfeta_{t|t-1}$ and covariance matrix $\bfP_{t|t-1}$ given as follows:
\begin{eqnarray} \label{eqn: one-step-ahead forecast}
\bfeta_{t|t-1} &=& \bfH_t \bfeta_{t-1|t-1}, \\
\bfP_{t|t-1} &=& \bfH_t \bfP_{t-1|t-1} \bfH_t' + \bfU_t,
\end{eqnarray}
where $\bfeta_{t-1|t-1}$ is the conditional mean and $\bfP_{t-1|t-1}$ is the conditional covariance matrix for the filtering distribution $[\bfeta_{t-1}|\bfZ_{1:t-1}]$.

The filtering distributions can be derived using Bayes' theorem: $[\bfeta_t \mid \bfZ_{1:t}] \propto [\bfZ_t \mid \bfeta_t] [\bfeta_t \mid \bfZ_{1:t-1}]$ and $[\bfdelta_t^P \mid \bfZ_{1:t}] \propto [\bfZ_t \mid \bfdelta_t^P] [\bfdelta_t^P \mid \bfZ_{1:t-1}]$. As we assume Gaussianity for these random vectors, these filtering distributions also follow multivariate normal distributions. The filtering algorithm is proceeded sequentially for $t=1, \ldots, T$:
\begin{eqnarray} \label{eqn: filtering algorithm}
\bfeta_{t|t} &=& \bfeta_{t|t-1} + \bfG_t (\bfZ_t -\bfX_t\bfbeta_t-\bfS_t\bfeta_{t|t-1}) \label{eqn: filter for eta}\\
\bfP_{t|t} &=& \bfP_{t|t-1} - \bfG_t \bfS_t \bfP_{t|t-1} \\
\bfdelta_{t|t}^P &=& \bfB_t^P \bfQ_t^{-1} \bfB_t'[\bfS_t\bfP_{t|t-1} \bfS_t' + \bfD_t^{-1}]^{-1} (\bfZ_t - \bfX_t\bfbeta_t - \bfS_t\bfeta_{t|t-1}) \label{eqn: filter for delta}\\
\bfR_{t|t}^P &=& \bfB_t^P\bfQ_t^{-1}\bfB_t^{P'}   
	- \bfB_t^P \bfQ_t^{-1} \bfB_t' [\bfS_t\bfP_{t|t-1} \bfS_t' + \bfD_t^{-1}]^{-1} \bfB_t\bfQ_t^{-1}\bfB_t^{P'}
\end{eqnarray}
where $\bfG_t\equiv \bfP_{t|t-1} \bfS_t' (\bfS_t\bfP_{t|t-1} \bfS_t' + \bfD_t^{-1})^{-1} $ is the $r\times n_t$ Kalman gain matrix. $\bfD_t \equiv (\bfB_t \bfQ_t^{-1} \bfB_t' + \bfV_t)^{-1}$ is an $n_t$-by-$n_t$ matrix. $\bfB_t^P$ is the basis matrix in the GGM component, with $(i,j)$-th element being one if the $i$-th prediction location is the $j$-th BAU, and zero otherwise.

The smoothing distribution $[\bfeta_t \mid \bfZ_{1:T}]$ can be written as 
\begin{eqnarray*}
[\bfeta_t \mid \bfZ_{1:T}] = \int [\bfeta_t\mid \bfeta_{t+1}, \bfZ_{1:T}] [\bfeta_{t+1} \mid \bfZ_{1:T}] \,d \bfeta_{t+1},
\end{eqnarray*}
where $[\bfeta_t\mid \bfeta_{t+1}, \bfZ_{1:T}] \propto [\bfeta_{t+1} \mid \bfeta_t] [\bfeta_t \mid \bfZ_{1:t}]$ is derived using Bayes' theorem. The first term on the right-hand side is just the evolution distribution, and the second term on the right-hand side is the filtering distribution. Similar formulas can be derived for the smoothing distribution $[\bfdelta_t^P \mid \bfZ_{1:T}]$. The detailed derivation of these formulas can be found in Chapter 8 of \cite{Cressie2011}, where derivation is given for more general dynamic spatio-temporal models. Then the smoothing algorithm proceeds backwards in time for $t=T-1, T-2, \ldots, 1$: 
\begin{eqnarray} \label{eqn: smoothing algorithm}
\bfeta_{t|T} &=& \bfeta_{t|t} + \bfJ_t(\bfeta_{t+1|T} - \bfeta_{t+1|t}) \label{eqn: smoother for eta} \\
\bfP_{t|T} &=& \bfP_{t|t} + \bfJ_t(\bfP_{t+1|T} - \bfP_{t+1|t}) \bfJ_t' \\
\bfdelta_{t|T}^P &=& \bfdelta_{t|t}^P + \bfM_t(\bfeta_{t+1|T} - \bfeta_{t+1|t}) \label{eqn: smoother for delta} \\
\bfR_{t|T}^P &=&  \bfR_{t|t}^P + \bfM_t(\bfP_{t+1|T} - \bfP_{t+1|t}) \bfM_t'
\end{eqnarray}
where 
%\begin{eqnarray*}
$\bfJ_t \equiv \bfP_{t|t} \bfH_{t+1}' \bfP_{t+1|t}^{-1}$ and 
$\bfM_t \equiv - \bfB_t^P \bfQ_t^{-1} \bfB_t' \bfG_t'\bfH_{t+1}' \bfP_{t+1|t}^{'-1}$.
%\bfM_t &\equiv& - \bfA_t^P \bfQ_t^{-1} \bfA_t' (\bfS_t\bfP_{t|t-1} \bfS_t' + \bfD_t^{-1})^{-1} \bfS_t \bfP_{t|t-1}' \bfH_{t+1}' \bfP_{t+1|t}^{'-1}.
%\end{eqnarray*}

 The smoothing distribution of the initial state $\bfeta_0$ has mean  
$ \bfeta_{0|T} = \bfeta_{0|0} + \bfJ_0(\bfeta_{1|T} - \bfeta_{1|t})$ and covariance matrix
  $\bfP_{0|T} = \bfP_{0|0} + \bfJ_0(\bfP_{1|T} - \bfP_{1|t}) \bfJ_{0}'$.

The lag-1 cross-covariance term $\bfP_{t,t-1|T} \equiv \text{cov}(\bfeta_t, \bfeta_{t-1} \mid\bfZ_{1:T})$ is given by 
\begin{eqnarray} 
\bfP_{T,T-1|T} &=& (\mathbf{I}_r - \bfG_T \bfS_T') \bfH_T \bfP_{T-1|T-1} \\
\bfP_{t,t-1|T} &=& \bfP_{t|t} \bfJ_{t-1}' + \bfJ_t (\bfP_{t+1,t|T} -  \bfP_{t|t}) \bfJ_{t-1}', \, t=T-1, T-2, \ldots, 1.
\end{eqnarray}

\subsection{Filtering Distribution for Hidden Process $Y_t(\cdot)$}
For $t=1, \ldots, T$, the optimal filter of $\bfY_{t}^P$ given the data $\bfZ_{1:t}$, denoted by $\bfY_{t|t}^P$, is 
\begin{eqnarray} \label{eqn: filtered predictor}
\bfY_{t|t}^P \equiv E(\bfY_t^P \mid \bfZ_{1:t}) = \bfX_t^P \bfbeta_t + \bfS_t^P \bfeta_{t|t} + \bfdelta_{t|t}^P,
\end{eqnarray}
where $\bfeta_{t|t}$ is given in Eq.~\eqref{eqn: filter for eta}, and $\bfdelta_{t|t}^P$ is given in Eq.~\eqref{eqn: filter for delta}. We call \eqref{eqn: filtered predictor} the \emph{Dynamic Fused Gaussian Process Filter (DFGPF)}. Its associated mean-squared-prediction-error covariance matrix is
\begin{eqnarray} \label{eqn: filter variance}
\boldsymbol \sigma^2_{t|t} \equiv E\{[\bfY^P - \bfY_{t|t}^P][\bfY^P - \bfY_{t|t}^P]'\}  
		       = \bfS_t^P\bfP_{t|t} \bfS_t^{P'} + \bfR_{t|t}^{P} + \bfS_t^P \bfC_{t|t} + (\bfS_t^P \bfC_{t|t})',  
\end{eqnarray}
where $\mathbf{C}_{t|t}\equiv \text{cov}(\bfeta_t, \bfdelta_t^P \mid \bfZ_{1:t}) = -\bfG_t\bfB_t\bfQ_t^{-1}\mathbf{B}_t^{P'}$. We call the square root of the diagonal elements in $\boldsymbol \sigma^2_{t|t}$ the DFGPF standard errors.

\subsection{Smoothing Distribution for Hidden Process $Y_t(\cdot)$}
For $t=1, \ldots, T-1$, the optimal smoother of $\bfY_t^P$ given the data $\bfZ_{1:T}$, denoted by $\bfY_{t|T}^P$, is 
\begin{eqnarray} \label{eqn: smoothed predictor}
\bfY_{t|T}^P \equiv E(\bfY_t^P \mid \bfZ_{1:T}) = \bfX_t^P \bfbeta_t + \bfS_t^P \bfeta_{t|T} + \bfdelta_{t|T}^P,
\end{eqnarray} 
where $\bfeta_{t|T}$ is given in Eq.~\eqref{eqn: smoother for eta}, and $\bfdelta_{t|T}^P$ is given in Eq.~\eqref{eqn: smoother for delta}. We call \eqref{eqn: smoothed predictor} the \emph{Dynamic Fused Gaussian Process Smoother (DFGPS)}. Its associated mean-squared-prediction-error covariance matrix is 
\begin{eqnarray}
\boldsymbol \sigma^2_{t|T} \equiv E\{[\bfY^P - \bfY_{t|T}^P][\bfY^P - \bfY_{t|T}^P]'\} 
	= \bfS_t^P\bfP_{t|T} \bfS_t^{P'} + \bfR_{t|T}^{P}   + \bfS_t^{P} \mathbf{C}_{t|T}  + (\bfS_t^{P} \mathbf{C}_{t|T})',
\end{eqnarray}
where  $\mathbf{C}_{t|T}\equiv \text{cov}(\bfeta_t, \bfdelta_t^P \mid \bfZ_{1:T})=  -\bfG_t\bfB_t\bfQ_t^{-1}\bfB_t^{P'} + \bfJ_t(\bfP_{t+1|T}-\bfP_{t+1|t})\bfM_t'$. We call the square root of the diagonal elements in $\boldsymbol \sigma^2_{t|T}$ the DFGPS standard errors.

\section{Maximum Likelihood Estimation via Stochastic EM Algorithm} \label{sec: SEM}
In what follows, we give a general derivation of the maximum likelihood estimation procedure for both filtering and smoothing methodology based on DFGP. Let $u$ be a generic time point. For the filtering-type estimator, the parameters will be estimated based on data $\bfZ_{1:u}$, where $u$ takes values from 2 to $T$, since data are collected from time point 1 to time point $T$. For the smoothing-type estimator, the parameters will be estimated based on the data $\bfZ_{1:u}$, where $u$ takes value $T$ only, since all the available data should be used in the smoothing-type methodology. To avoid identifiability issues, we assume that the propagation matrices $\{\bfH_t: t=1, \ldots, T\}$ and innovation matrices $\{\bfU_t: t=1, \ldots, T\}$ are time-invariant with common propagation matrix $\bfH$ and common innovation matrix $\bfU$. Let $\bftheta \equiv \{ \bfbeta_1, \ldots, \bfbeta_u, \bfK_0, \bfH, \bfU, \tau^2_1, \ldots, \tau^2_u, \gamma_1, \ldots, \gamma_u\} \cup \{\sigma^2_{\epsilon_t, (k)}: t=1, \ldots, u; k=1, \ldots, k_0\}$ be a collection of model parameters up to time $u$. If the goal is to make filtering-type predictions, the letter $u$ denotes the current time at which predictions will be made, and parameters are estimated based on data up to current time $u$. If the goal is to make smoothing-type predictions, the letter $u$ denotes the time at which latest data are observed at time $T$. Smoothing-type predictions will be made at time $t=1, \ldots, T-1$. In what follows, we give an efficient parameter estimation procedure to ensure the scalability of the DFGP methodology for massive datasets.

\subsection{Likelihood Function}
Let $\bfalpha_t = \bfZ_t - \bfX_t\bfbeta_t - \bfS_t \bfeta_{t|t-1}$ be innovations for $t=1, \ldots, u$. These innovations are independently and normally distributed with mean zero and covariance matrix $\bfSigma_{t|t-1}=\bfS_t\bfP_{t|t-1}\bfS_t' + \bfD_t^{-1}$. Then, up to a constant, the negative twice marginal log-likelihood function is 
\begin{eqnarray} \label{eqn: loglikelihood}
-2\ln L(\bftheta) = -2 f(\bfalpha_1, \ldots, \bfalpha_u| \bftheta) = \sum_{t=1}^u \ln |\bfSigma_{t|t-1}| + \sum_{t=1}^u \bfalpha_t(\bftheta)' \bfSigma_{t|t-1}(\bftheta)^{-1} \bfalpha_t(\bftheta),
\end{eqnarray}
where $\bftheta$ denotes model parameters. The inverse and log-determinant of $\bfSigma_{t|t-1}$ can be calculated as follows:
\begin{eqnarray} \label{eqn: matrix inversion}
\bfSigma_{t|t-1}^{-1} &=& \bfD_t - \bfD_t \bfS_t (\bfP_{t|t-1}^{-1} + \bfS_t' \bfD_t \bfS_t)^{-1} \bfS_t' \bfD_t \\
\ln |\bfSigma_{t|t-1}| &=& \ln |\bfP_{t|t-1}^{-1} + \bfS_t' \bfD_t \bfS_t| + \ln |\bfP_{t|t-1}| + \ln |\bfD_t^{-1}|,
\end{eqnarray}
where $\bfD_t=\bfV_t^{-1} -\bfV_t^{-1} \bfB_t (\bfQ_t + \bfB_t' \bfV_t^{-1} \bfB_t)^{-1} \bfB_t' \bfV_t^{-1}$ and $\ln |\bfD_t^{-1} | = \ln |\bfQ_t+\bfB_t'\bfV_t^{-1} \bfB_t| - \ln |\bfQ_t^{-1} | + \ln |\bfV_t|$.
To evaluate the negative twice marginal log-likelihood function in Eq.~\eqref{eqn: loglikelihood}, solving linear systems involving $\bfSigma_{t|t-1}$ is required. To solve $\bfSigma_{t|t-1}^{-1} \bfT$ for an $n_t$-dimensional vector or an $n_t$-by-$r$ matrix $\bfT$, one has to solve linear systems involving $r$-by-$r$ matrices and $N$-by-$N$ sparse matrix $\bfQ_t + \bfB_t' \bfV_t^{-1} \bfB_t$. The former requires $O(r^3)$ computational cost. If the sparse matrix $\bfQ_t + \bfB_t' \bfV_t^{-1} \bfB_t$ has bandwidth $p_0$ after appropriate reordering, the computational cost of the Cholesky decomposition for $\bfQ_t + \bfB_t' \bfV_t^{-1} \bfB_t$ is $O(N(p_0^2+3p_0))$. Linear systems involving $\bfQ_t + \bfB_t' \bfV_t^{-1} \bfB_t$ can be solved efficiently. The Cholesky decomposition of $\bfQ$ can be solved very efficiently with $O(N^{1.5})$ computational cost \cite[see][]{Ma2017}.

\subsection{Stochastic EM Algorithm}
The expectation-maximization (EM) algorithm introduced by \cite{Dempster1977} is a powerful method to solve maximum likelihood estimation problems iteratively, where the E-step is calculated exactly and then is followed by an M-step in each iteration. Instead of computing the conditional expectation exactly in the E-step, one can employ Monte Carlo algorithms to generate samples from the conditional distribution, and then replace the conditional expectation with an average of corresponding quantities evaluated at these samples. This estimation method is called the Monte Carlo EM algorithm \citep{Wei1990}. The success of the Monte Carlo EM algorithm relies on sufficiently large samples to approximate the E-step conditional expectations. A closely related modification of the EM algorithm described in \cite{Celeux1985} is known as the stochastic EM (SEM) algorithm, which substitutes the E-step with a single corresponding quantity evaluated with one random sample from the conditional distribution. The SEM algorithm is generally less computationally expensive than the Monte Carlo EM algorithm. It is robust with initial values and converges to a stationary distribution \cite[see][for details]{Celeux1985, Diebolt1996, Nielsen2000}. However, when computing resources are less constrained, one can employ the Monte Carlo EM algorithm estimate parameters in the DFGP methodology. In what follows, we give the SEM procedure to estimate parameters in the DFGP methodology.

In the EM algorithm of the DFGP model, we treat $\bfeta_{0:u}$ and $\bfxi_{1:u}$ as ``missing data''. Up to a constant, the negative twice complete-data log-likelihood is 
\begin{eqnarray} \label{eqn: complete loglikelihood} 
-2\ln L_c(\bftheta) &=& -2\ln f(\bfZ_{1:u}, \bfeta_{0:u}, \bfxi_{1:u} | \bftheta) \\ \nonumber
&=& \sum_{t=1}^u \left\{(\bfZ_t - \bfX_t \bfbeta_t - \bfS_t \bfeta_t - \bfB_t \bfxi_t)' \bfV_t^{-1} (\bfZ_t - \bfX_t \bfbeta_t - \bfS_t \bfeta_t - \bfB_t \bfxi_t) + \ln |\bfV_t|\right\} \\ \nonumber
&+& \sum_{t=1}^u \left\{(\bfeta_t - \bfH\bfeta_{t-1})' \bfU^{-1} (\bfeta_t - \bfH \bfeta_{t-1}) + \ln |\bfU|\right\} \\ \nonumber
&+& \ln |\bfK_0| + \bfeta_0' \bfK_0^{-1} \bfeta_0 + \sum_{t=1}^u\left\{ \bfxi_t'\bfQ_t \bfxi_t - \ln|\bfQ_t|\right\}. 
\end{eqnarray}
Given the negative twice complete-data log-likelihood function in Eq.~\eqref{eqn: complete loglikelihood}, we now derive the $Q$-function in the EM algorithm first, and then present the derivation for the SEM algorithm. Consider the $(\ell+1)$th iteration in the EM algorithm. The E-step is to find conditional expectation of the complete-data log-likelihood for $\bftheta=\bftheta^{[\ell]}$ with respect to missing data, i.e., $Q(\bftheta; \bftheta^{[\ell]}) := E_{\bftheta^{[\ell]}} [ -2\ln L_c(\bftheta) \mid \bfZ_{1:u}]$. 
In what follows, the following notations are used: $\bfeta_{t|u}^{[\ell]} = E_{\bftheta^{[\ell]}}(\bfeta_t\mid \bfZ_{1:u}), \bfdelta_{t|u}^{[\ell]} = E_{\bftheta^{[\ell]}} (\bfdelta_t\mid \bfZ_{1:u})$, $\bfP_{t|u}^{[l]} = \text{var}(\bfeta_t\mid \bfZ_{1:u}, \bftheta^{[\ell]}), \bfR_{t|u}^{[\ell]} = \text{var} (\bfdelta_t \mid \bfZ_{1:u}, \bftheta^{[l]})$, $\bfP_{t,t-1|u}^{[\ell]} = \text{cov}(\bfeta_t, \bfeta_{t-1} \mid \bfZ_{1:u}, \bftheta^{[\ell]})$ for all $t$. 
In the EM algorithm, the actual prediction location $\mathcal{R}^P$ is the same as the observed locations, $\mathcal{R}^O$, for $t=1, \ldots, u$. We also define the quantities $\bfK_t^{[\ell+1]} \equiv \bfP_{t|u}^{[\ell]} + \bfeta_{t|u}^{[\ell]}\bfeta_{t|u}^{[\ell]'}$ and $\bfL_t^{[\ell+1]} \equiv \bfP_{t,t-1|u}^{[\ell]} + \bfeta_{t|u}^{[\ell]} \bfeta_{t-1|u}^{[\ell]'}$. 

In the EM algorithm, these conditional expectations are computed exactly; while in the SEM algorithm, they are replaced by the complete log-likelihood function evaluated with a single sample from the posterior distribution $[\bfeta_t, \bfxi_t \mid \bfZ_{1:u}]$. To ensure the positive definiteness of matrices $\bfK_0^{[\ell]}$ and $\bfU_0^{[\ell]}$, we compute the conditional expectations involving them exactly. We only use samples from $[\bfeta_t, \bfxi_t \mid \bfZ_{1:u}, \bftheta^{[\ell]}]$ to approximate the conditional expectations involving in $\bfxi_t$. That is, we compute conditional expectations involving $\bfeta_t$ explicitly, and we define $\bfeta_{t|u}^{[\ell]} = E_{\bftheta^{[\ell]}}(\bfeta_t \mid \bfZ_{1:u})$. For expectations involving $\bfxi_t$, we generate a sample from $[ \bfeta_t, \bfxi_t \mid \bfZ_{1:u}, \bftheta^{[\ell]}]$, and use $\bfxi_{t|u}^{[\ell]}$ to denote a sample for $\bfxi_t$ from the distribution $[ \bfeta_t, \bfxi_t \mid \bfZ_{1:u}, \bftheta^{[\ell]}]$. To generate a sample from the distribution $[ \bfeta_t, \bfxi_t \mid \bfZ_{1:u}, \bftheta^{[\ell]}]$, we use the conditional simulation strategy in geostatistics to allow fast sampling procedure \citep[for details, see][Sec. 3.6.2]{Cressie1993}. The negative twice $Q_{SEM}(\cdot; \cdot)$ function in the SEM algorithm is 
\begin{eqnarray*}
-2Q_{SEM}(\bftheta; \bftheta^{[\ell]}) %&=& E_{\bftheta^{[\ell]}} [ -2\ln L_c(\bftheta) | \bfZ_{1:T}] \\
&\equiv& \sum_{t=1}^u \left\{(\bfZ_t - \bfX_t \bfbeta_t - \bfS_t \bfeta_{t|u}^{[\ell]} - \bfB_t \bfxi_{t|u}^{[\ell]})' \bfV_t^{-1} (\bfZ_t - \bfX_t \bfbeta_t - \bfS_t \bfeta_{t|u}^{[\ell]} - \bfB_t \bfxi_{t|u}^{[\ell]}) \right\}\\
&+& u\ln |\bfU| + \sum_{t=1}^u tr\left\{ \bfU^{-1} (\bfK_t^{[\ell+1]} - \bfH \bfL_t^{[\ell+1]'} - \bfL_t^{[\ell+1]}\bfH' + \bfH\bfK_{t-1}^{[\ell+1]}\bfH') \right\} \\
&+& \ln |\bfK_0| + tr(\bfK_0 \bfK_0^{[\ell+1]}) + \sum_{t=1}^u \left\{ \bfxi_{t|u}^{[\ell]'} \bfQ_t \bfxi_{t|u}^{[\ell]} - \ln|\bfQ_t| + \ln |\bfV_t| \right\},
\end{eqnarray*}
where $\bfxi_{t|u}^{[\ell]}$ is the sub-vector of a random sample from $[\bfeta_t, \bfxi_t \mid \bfZ_{1:u}, \bftheta^{[\ell]}]$ corresponding to $\bfxi_t$. We compute the expectations involving $\bfU$ and $\bfK_0$ exactly, since this will give a desirable property in their updating formulas that both of them are guaranteed to be positive definite in each iteration of the SEM algorithm.   

In the M-step, this $Q_{SEM}$ function is maximized with respect to parameters $\bftheta$, yielding the following formulas to update parameters for $t=1, \ldots, u$:
\begin{eqnarray*} \label{eqn: parameter updates}
\bfbeta_t^{[\ell+1]} &=& (\bfX_t' \bfV_t \bfX_t)^{-1} \bfX_t' \bfV_t^{-1} (\bfZ_t - \bfS_t \bfeta_{t|u}^{[\ell]} - \bfB_t \bfxi_{t|u}^{[\ell]}), \\ \nonumber
\sigma_{\epsilon_t, (k)}^{2 [\ell+1]} &=& (\bfZ_t^{(k)} - \bfX_t^{(k)}\bfbeta_{t}^{[\ell+1]}- \bfS_t^{(k)} \bfeta_{t|u}^{[\ell]} - \bfB_t^{(k)} \bfxi_{t|u}^{[\ell]})' \bfV_{\epsilon_t, (k)}^{-1} \\
&\cdot& (\bfZ_t^{(k)} - \bfX_t^{(k)}\bfbeta_{t}^{[\ell+1]}- \bfS_t^{(k)} \bfeta_{t|u}^{[\ell]} - \bfB_t^{(k)} \bfxi_{t|u}^{[\ell]})
 \Big/ n_t^{(k)}, \\ 
\bfK_0^{[\ell+1]} &=& \bfP_{0|u}^{[\ell]} + \bfeta_{0|u}^{[\ell]} \bfeta_{0|u}^{[\ell]'}, \\
\bfH^{[\ell+1]} &=& \left(\sum_{t=1}^u \bfL_t^{[\ell+1]}\right) \left(\sum_{t=0}^{u-1} \bfK_t^{[\ell+1]}\right)^{-1}, \\
\bfU^{[\ell+1]} &=& \left(\sum_{t=1}^u \bfK_t^{[\ell+1]} - \bfH^{[\ell+1]} \sum_{t=1}^u \bfL_t^{[\ell+1]'}\right) \Big/u, \\ 
\tau_t^{2[\ell+1]} &=& \bfxi_{t|u}^{[\ell]'} (\mathbf{I} - \gamma_t^{[\ell]}\bfW) \bfxi_{t|u}^{[\ell]} \Big/ N, 
\end{eqnarray*}
where the parameters $\beta_t, \sigma^2_{\epsilon_t, (k)}, \mathbf{K}_0, \mathbf{H}, \mathbf{U}$, and $\tau^2_t$ have closed-form updates. When the M-step is carried out w.r.t.~$\{\gamma_t: t=1, \ldots, u\}$, there is no explicit formula. However, it is straightforward to show that we can minimize the following function w.r.t.~these parameters:
\begin{eqnarray*} 
g(\gamma_1, \ldots, \gamma_u) &=& \sum_{t=1}^u ( -\gamma_t\bfxi_{t|u}^{[\ell]'} \bfW \bfxi_{t|u}^{[\ell]}/\tau_t^{2[\ell]} - \ln|\mathbf{I} - \gamma_t\bfW|).
\end{eqnarray*}
To solve this nonlinear optimization problem, we use the interior-point method \citep[e.g.,][]{Byrd1999}. Notice that we can estimate all the parameters including the nugget variance parameters $\{ \sigma^2_{\epsilon_t, (k)}: t=1, \ldots, u; k=1, \ldots, k_0.\}$ in the SEM algorithm.

The SEM algorithm starts with certain initial values for parameters $\bftheta$, and then these parameters are updated iteratively at each iteration. The initial values should be tuned to achieve better convergence results. Here, we give some practical suggestions. The initial values of regression coefficients can be set as the ordinary least square estimates. The initial values for $\{ \sigma^2_{\epsilon_t, (k)}: t=1, \ldots, u; k=1, \ldots, k_0.\}$ can be set as the parameter estimates by fitting empirical semivariograms near origin \citep[e.g.,][]{Kang2010}. The initial values for $\bfK_0$ and $\bfU$ can be set to the $r$-by-$r$ positive definite matrices with diagonal entries adjusted by the empirical variance of the data values $\{\bfZ_t: t=1, \ldots, u\}$. The initial values for $\tau^2_t$ can be set as a small portion (say 0.01) of the empirical variance of the data values $\{\bfZ_t: t=1, \ldots, u\}$. The initial values for $\bfH$ can be set as an identity matrix. After the initial values are specified, the SEM algorithm will update them in each iteration. The formulas to update all these parameters reveal that the positive definiteness of matrix $\bfK_0$ and $\bfU$ is guaranteed. To check the convergence of the SEM algorithm, we can monitor the change or relative change of log-likelihood function \eqref{eqn: loglikelihood} for a sufficient number of consecutive iterations. In addition, one can also monitoring the different of parameter values, i.e., $\| \bftheta^{[\ell+1]} - \bftheta^{[\ell]}\|$. 

Although the procedure to obtain standard errors for the parameter estimates is not discussed here, these standard errors can be obtained in a certain way. For instance, one might use the bootstrap-sampling technique described in \cite{Stoffer1991} to compute the standard errors for these parameters estimates. Its detailed discussion is beyond the scope of this work. As in each iteration of the SEM algorithm, solving linear systems involving $r$-by-$r$ matrices and $N$-by-$N$ sparse matrices is required. We also need to carry out numerical optimization to update $\{\gamma_t: t=1, \ldots, u\}$, which requires the Cholesky decomposition for sparse matrices $\{\bfQ_t: t=1, \ldots, u\}$. These computations can be done efficiently but still require considerate amount of computing time. 

When data are sparse, we can impose the time-invariant assumption for the nugget variance parameters $\{\sigma^2_{\epsilon_t, (k)}: t=1, \ldots, u; k=1, \ldots, k_0.\}$ so that stable estimates for these parameters can be obtained. Suppose that $\sigma^2_{\epsilon, (k)} \equiv \sigma^2_{\epsilon_1, (k)} = \ldots, = \sigma^2_{\epsilon_u, (k)}$. The formulas to update these nugget variance parameters are: 
\begin{eqnarray*}
 \sigma_{\epsilon, (k)}^{2 [\ell+1]} &=& \sum_{t=1}^u  (\bfZ_t^{(k)} - \bfX_t^{(k)}\bfbeta_{t}^{[\ell+1]}- \bfS_t^{(k)} \bfeta_{t|u}^{[\ell]} - \bfB_t^{(k)} \bfxi_{t|u}^{[\ell]})' \bfV_{\epsilon_t, (k)}^{-1} \\
&\cdot &  (\bfZ_t^{(k)} - \bfX_t^{(k)}\bfbeta_{t}^{[\ell+1]}- \bfS_t^{(k)} \bfeta_{t|u}^{[\ell]} - \bfB_t^{(k)} \bfxi_{t|u}^{[\ell]})  \bigg/ \left(\sum_{t=1}^u n_t^{(k)} \right).
\end{eqnarray*}
When data are available over a large number of time steps, i.e., $u$ is large, we can relax the time-invariant assumption for propagation matrices $\{\bfH_t: t=1, \ldots, u\}$, and innovation matrices $\{\bfU_t: t=1, \ldots, u\}$. Suppose that the propagation matrix and innovation matrix are constant over $b_0$ block time periods. Let $T_0\equiv 0$ and $T_{b_0}\equiv u$.  We further assume that $\bfH_{1}=\ldots =\bfH_{T_1} \neq \bfH_{T_1+1}  = \ldots =\bfH_{T_2} \neq \ldots  =\bfH_{T_{b_0}}$ and $\bfU_{1}=\ldots =\bfU_{T_1} \neq \bfU_{T_1+1}  = \ldots = \bfU_{T_2} \neq \ldots = \bfU_{T_{b_0}}$. For $b=1, \ldots, b_0$, the formulas to update $\bfH_{T_b}$ and $\bfU_{T_b}$ are 
\begin{eqnarray*}
\bfH_{T_b}^{[\ell+1]} &=& \left(\sum_{t=T_{b-1}+1}^{T_{b}} \bfL_t^{[\ell+1]}\right) \left(\sum_{t=T_{b-1}+1}^{T_{b}-1} \bfK_t^{[\ell+1]}\right)^{-1}, \\
\bfU_{T_b}^{[\ell+1]} &=& \left(\sum_{t=T_{b-1}+1}^{T_{b}} \bfK_t^{[\ell+1]} - \bfH^{[\ell+1]}_t \sum_{t=T_{b-1}+1}^{T_b} \bfL_t^{[\ell+1]'}\right) \bigg/\left(T_b - T_{b-1}\right).
\end{eqnarray*}

%\section{Numerical Illustration}

%\section{Simulation Study}

\section{Results} \label{sec: results}
In what follows, we demonstrate DFGP methodology based on two SST datasets from MODIS and AMSR-E instruments. The tropical Pacific region is covered by $N=1,260,864$ grid cells at 9km resolution, which are used to define BAUs in this study. The CAR structure in the DFGP model is constructed based on first-order neighborhood structure on these BAUs. These two SST datasets have been bias-corrected to ensure that the mean zero assumption in the measurement-error process is valid in the DFGP methodology. Exploratory analysis suggests that the covariates $\bfX_t(\cdot) = [1, \text{latitude}(\cdot)$, $\text{latitude}(\cdot)^2]$ be included in the trend component for all time points. In Section~\ref{subsec: filtering CV} and Section~\ref{subsec: smoothing CV}, we demonstrate filtering and smoothing procedure of the DFGP methodology, respectively. Comparison of DFGP with with other existing methods is also made. In the following numerical illustrations, to assess the predictive performance for each method, we use the root-mean-sqaured-prediction error (RMSPE) to evaluate the accuracy of the predictions at held-out locations. We also use the continuous-rank-probability score \citep[CRPS;][]{Gneiting2007} to evaluate the quality of predictions for held-out locations, where small value of the CRPS indicates better prediction. 

\subsection{Cross-Validation Study in the Filtering Methodology} \label{subsec: filtering CV}
We first carry out cross-validation to evaluate the filtering-type DFGP methodology and compare it with the spatio-temporal data fusion model in \cite{Nguyen2014}. In what follows, we refer to the spatio-temporal data fusion model in \cite{Nguyen2014} as \emph{Fixed Rank Filtering (FRF)}, and we refer to our filtering procedure of the DFGP methodology as \emph{DFGPF}. In FRF, we consider 99 equally-spaced basis functions at three different resolutions and 181 equally-space basis functions at four different resolutions with previous 99 basis functions included. As an additional comparison, we also implement a local kriging approach \citep[e.g.,][]{Haas1990, Vecchia1988, Kuusela2018}, which makes predictions based on observations in a moving window. This approach has been widely used in practice to make spatial/spatio-temporal predictions due to its computational efficiency and the usage of spatial/spatio-temporal varying covariance functions in a local moving window. In our implementation of the local kriging approach, we fit a spatio-temporal exponential covariance function model with its parameters estimated by the maximum likelihood method in a spatio-temporal moving window such that it contains 500 nearest observations. The exponential covariance function is chosen to be anisotropic in space and time, i.e., $C(h, u) = \sigma^2\exp\{-\sqrt{\frac{h^2}{\phi_s^2} + \frac{u^2}{\phi_t^2}  } \} + \sigma^2_{\epsilon}I(h=0, u=0)$, where $h$ is the chordal distance in space and $u$ is the Euclidean distance in time. $\phi_s$ and $\phi_t$ are range parameters in space and time, $\sigma^2$ is the partial sill, and $\sigma^2_{\epsilon}$ is the nugget variance. This approach will be referred to as Local Kriging hereafter. Notice that when implementing the local kriging approach, we assume that each observation is associated with the centroid of the grid cell, and we ignore the resolution differences among each dataset. In the filtering context, the nearest observations are selected based on all training observations from $t=1, \ldots, 8$. 

\begin{figure}[htbp]
\begin{center}
\makebox[\textwidth][c]{ \includegraphics[width=1.0\textwidth, height=0.3\textheight]{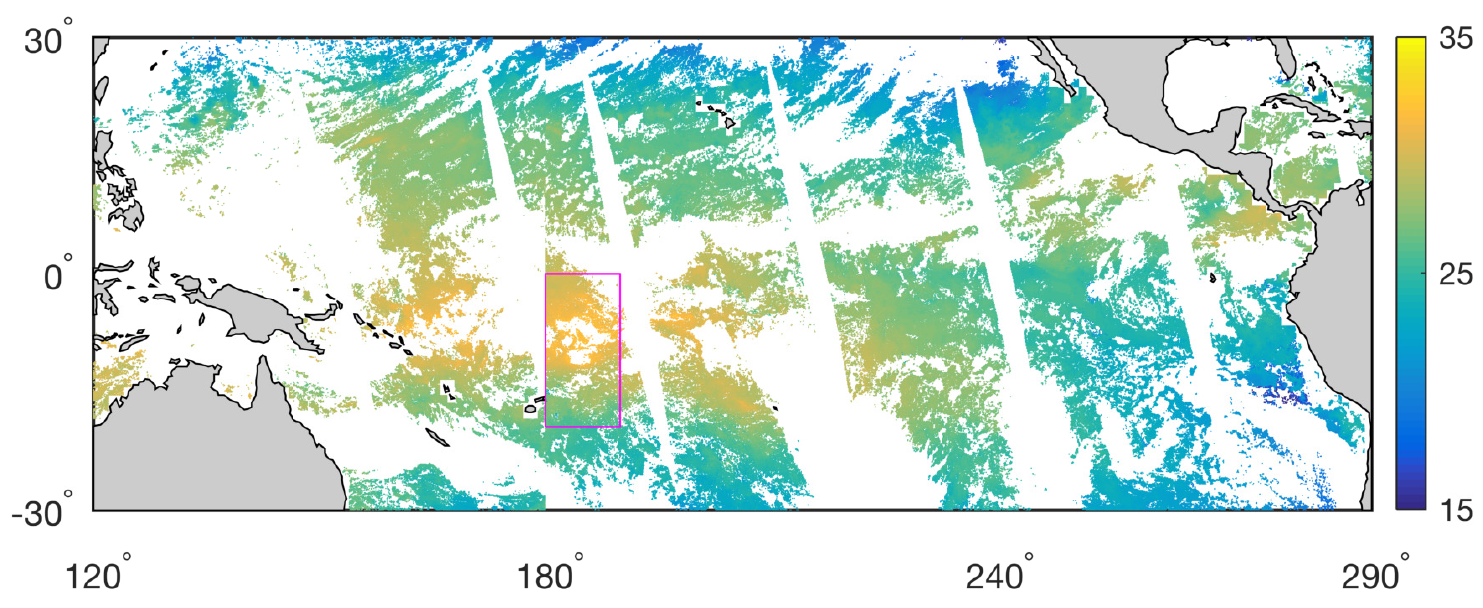}}
\caption{MODIS SST data are held out in the rectangular region on January 8, 2010. The delineated rectangular region is the held-out contiguous region to test long-range prediction skills.}
\label{fig: block region SST}
\end{center}
\end{figure}

To set up the cross-validation exercise, we hold out MODIS SST in the block region between longitude $180^{\circ}$ and $190^{\circ}$ and between latitude $-20^{\circ}$ and $0^{\circ}$ for time point $t=2, \ldots, 8$; see Figure~\ref{fig: block region SST} for an example of the held-out region on January 8, 2018. This large contiguous region is used to test long-range prediction skills that are often required to make predictions for remote sensing data, because remote sensing data often have missing latitude bands when polar-orbiting satellite instruments collect the data around the Earth. In addition, we also hold out randomly sampled 10\% of remaining MODIS SST observations for time point $t=2,  \ldots, 8$. These randomly held-out prediction locations will be used to test the short-range prediction skills.  As our focus is to compare filtering-type predictions among Local Kriging, FRF and DFGPF, we only use the data up to time $t$ to estimate parameters and to make filtering predictions at held-out locations at time $t$, where $t=2,  \ldots, 8$. Notice that we do not hold out observations at time $t=1$, since the filtering-type DFGP methodology reduces to the spatial-only FGP methodology, which has been studied in \cite{Ma2017} under various numerical examples.

\begin{figure}[htbp]
\begin{center}
\makebox[\textwidth][c]{ \includegraphics[width=1.0\textwidth, height=0.2\textheight]{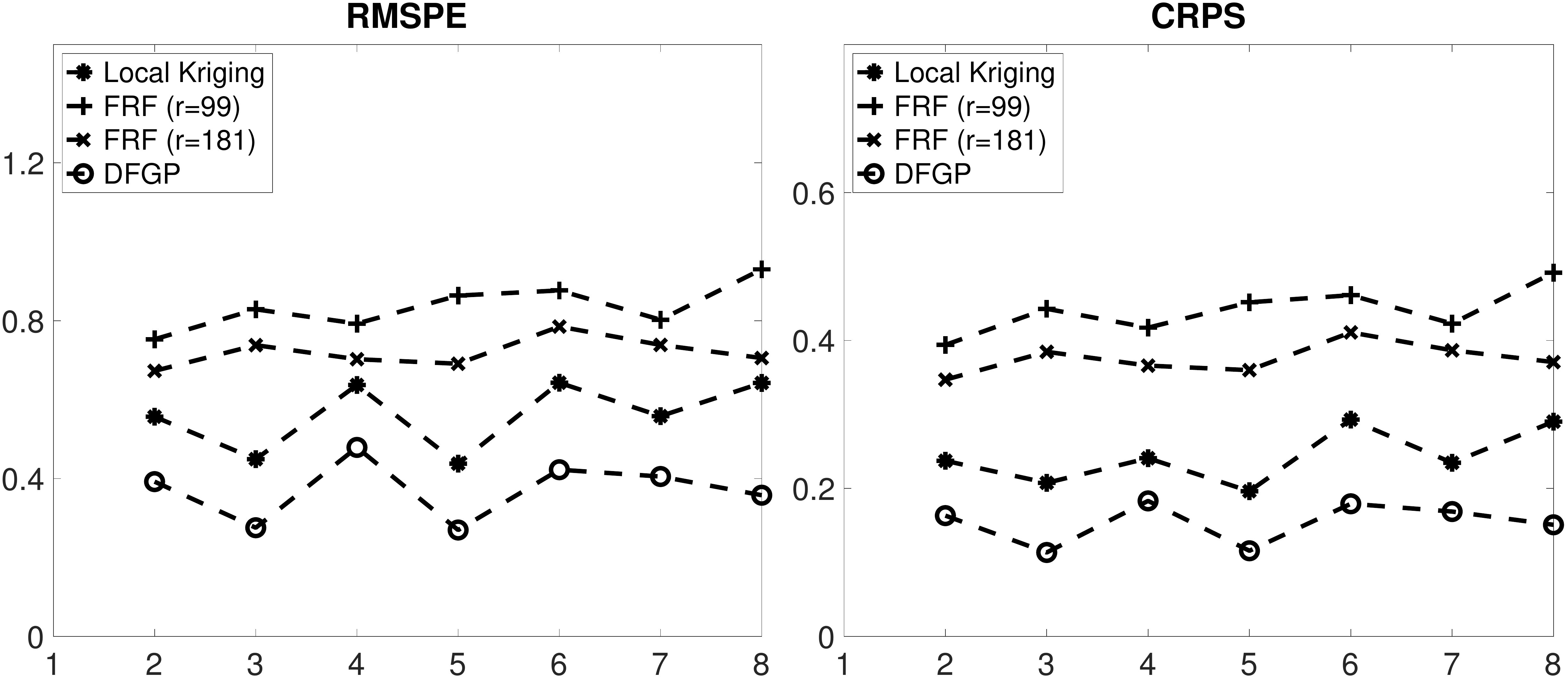}}
\caption{Numerical measures for predictions over all held-out locations based on Local Kriging, FRF, and DFGPF at $t=2, \ldots, 8$. The figure shows the RMSPEs at the left panel and CRPSs at the right panel for these methods, respectively. The asterisk represents the numerical measures based on Local Kriging; the plus sign represents the numerical measures based on FRF with $r=99$ basis functions; the cross sign represents the numerical measures based on FRF with $r=181$ basis functions; the circle sign represents the numerical measures based on DFGPF.}
\label{fig: numerical comparison at different time points}
\end{center}
\end{figure}

Figure~\ref{fig: numerical comparison at different time points} shows the RMSPE and CRPS for FRF, DFGPF and Local Kriging at time point $t=2,  \ldots, 8$. As we can see, DFGPF performs the best among all the three methods in terms of RMSPE and CRPS at all time points. Even though the number of basis functions in the low-rank component of FRF is almost twice as that in DFGPF, DFGPF still outperforms FRF in terms of RMSPE and CRPS at all time points, since the DFGPF model incorporates a more flexible covariance function than that in the FRF model. The Local Kriging approach gives better predictions than FRF. However, FRF may perform better than Local Kriging using adaptive basis functions; readers are referred to \cite{Ma2018downscaling} for details. In addition, we also tried in increase more basis functions, but this will pose numerical instabilities due to different contiguous regions at different time points. Basis function selection for such a spatio-temporal model is still an open problem. Methods in \cite{Tzeng2017} and \cite{Ma2018downscaling} can be extended to the spatio-temproral context to tackle this problem. This will be left for future research. DFGP outperforms Local Kriging at all tested time points in terms of RMSPE and CRPS. The reasons are as follows. First, Local Kriging fails to provide good predictions especially in large contiguous missing region, since it only uses local information without borrowing strength from distant observations. Second, Local Kriging uses a stationary exponential covariance function, which may not be flexible enough to capture nonstationary behavior of the underlying geophysical process. Third, Local Kriging does not address the change-of-support problem. It is shown in \cite{Ma2018downscaling} that ignoring the change-of-support problem can lead to unfavorable statistical inferential results. 

Table~\ref{table: CV_SST} shows the average of numerical measure such as RMSPE and CRPS across all the seven tested time points as well as the total computing time (in hours) for parameter estimation and prediction on a 10-core machine with 20GB memory and Intel Xeon E5-2680 central processing unit. We see that DFGPF outperforms Local Kriging and FRF in terms of RMSPE and CRPS. For computing time, FRF is fastest, since FRF only needs to invert $r$-by-$r$ matrix and diagonal matrix. When $r$ is very small (e.g., $r=99, 181$), its computation can be very fast. Local Kriging is second fastest, since it only needs to solve small (e.g., 500-by-500) linear systems for every prediction location, and the parallel computing environment can be employed to facilitate computations. 
DFGPF requires about 4 to 9 times, dependent on the number of basis functions in FRF, more computing time than FRF, since DFGPF not only needs to invert $r$-by-$r$ matrices but also needs to solve sparse linear systems for $N$-by-$N$ sparse matrices. Even though DFGPF requires more computing time, it can give very good predictive performance in a reasonable amount of time, since DFGPF is able to process a 8-day dataset with about 3.7 million observations in a time much less than one week.

%% block region

\subsection{Cross-Validation Study in the Smoothing Methodology} \label{subsec: smoothing CV}
DFGP not only allows filtering-type predictions, but also allows smoothing-type predictions. To demonstrate the smoothing methodology of DFGP, we compare it with the spatio-temporal data fusion model in \cite{Nguyen2014}. In what follows, we refer to our smoothing procedure of the DFGP methodology as \emph{DFGPS}, and refer to the spatio-temporal data fusion model in \cite{Nguyen2014} as \emph{Fixed Rank Smoothing (FRS)}. Notice that the methodology in FRS appeared first in \cite{Katzfuss2011} for a single data source. \cite{Nguyen2014} generalize this approach for multiple data sources. Similar to Section~\ref{subsec: filtering CV}, FRS is implemented with 99 equally-spaced basis functions at three different resolutions and 181 equally-spaced basis functions at four different resolutions. DFGPS is implemented with 99 basis functions that are the same as in FRS. As an additional comparison, we also implement Local Kriging. In the smoothing-type predictions, the nearest observations in Local Kriging are also selected based on all observations from January 1 to January 8, 2010. In the implementation of Local Kriging, we also chose 500 nearest observations. We also tried to increase the number of nearest observations, but improvement of prediction based on Local Kriging with more nearest observations is negligible and more computing resources are required in terms of computer memory and computing time.

To setup the cross-validation exercise, we hold out MODIS SST in the block region between longitude $180^{\circ}$ and $190^{\circ}$ and between latitude $-20^{\circ}$ and $0^{\circ}$ for time point $t=2, \ldots, 8$ as in Section~\ref{subsec: filtering CV}. Then 10\% of remaining MODIS SST are randomly held out to test the short-range prediction skills. Unlike the filtering context, the smoothing prediction of the process $Y_t(\cdot)$ for these held-out locations are obtained based on all remaining observations from both MODIS and AMSR-E instruments. Notice that the total number of observations that are predicted are much more in the smoothing context than those in the filtering context, since in the filtering context, we only evaluate prediction skills for $Y_t(\cdot)$ at current time $u$ based on data $\bfZ_{1:u}$, where $u=2, \ldots, 8$.

\begin{figure}[htbp]
\begin{center}
\makebox[\textwidth][c]{ \includegraphics[width=1.0\textwidth, height=0.2\textheight]{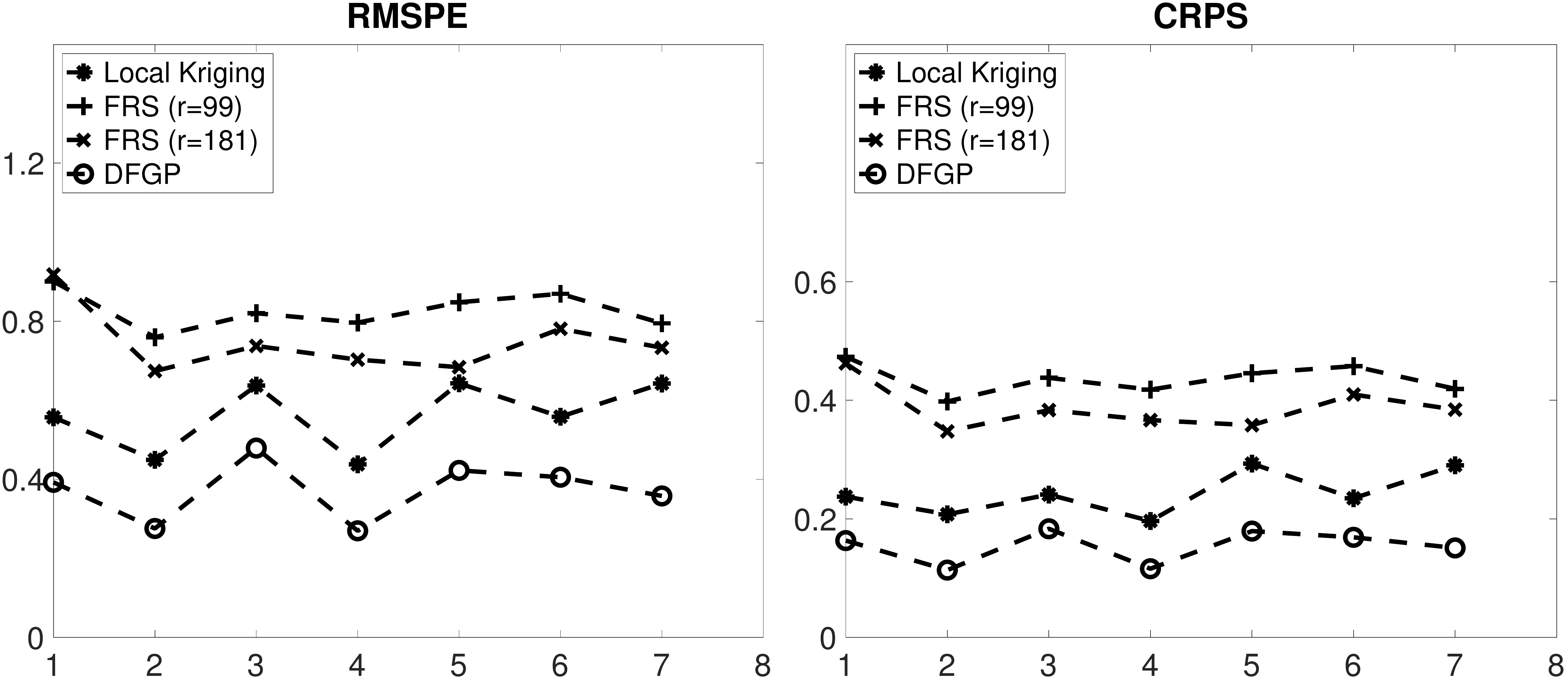}}
\caption{Numerical measures for predictions over all held-out locations based on Local Kriging, FRS, and DFGPS at $t=1, \ldots, 7$. The figure shows the RMSPEs at the left panel and CRPSs at the right panel for these methods, respectively. The asterisk represents the numerical measures based on Local Kriging; the plus sign represents the numerical measures based on FRS with $r=99$ basis functions; the cross sign represents the numerical measures based on FRS with $r=181$ basis functions; the circle sign represents the numerical measures based on DFGPS.}
\label{fig: numerical comparison in smoothing}
\end{center}
\end{figure}

Figure~\ref{fig: numerical comparison in smoothing} shows the RMSPE and CRPS for held-out locations at time point $t=1, \ldots, 7$ based on FRS, DFGPS, Local Kriging. As we can see, the DFGPS model performs better than all the other models in terms of RMSPE and CRPS at all time points. This suggests that our proposed DFGP methodology can provide good smoothing-type predictions. Even though the number of basis functions in FRS is twice as that in the low-rank component of DFGPS, FRS still cannot outperform DFGPS. This is consistent with findings in the spatial-only context in \cite{Ma2017} and the filtering context in Section~\ref{subsec: filtering CV}. This is because the CAR structure introduces a spatial dependence structure that can capture the unexplained variation by the low-rank component. We do not further increase the number of basis functions, since we encountered numerical instability due to large portions of contiguous missing regions. The optimal selection of spatial basis functions in a spatio-temporal context is very challenging especially when data have different patterns of large contiguous missing region over time. DFGPS performs better than Local Kriging, which is expected with the same reasons given in Section~\ref{subsec: smoothing CV}.

Table~\ref{table: smoothing CV_SST} shows the average of RMSPE and CRPS across all seven time points as well as the total computing time (in hours) for parameter estimation and prediction on a 10-core machine with 20GB memory and Intel Xeon E5-2680 central processing unit. We see that DFGPS outperforms Local Kriging and FRS in terms of RMSPE and and CRPS. FRS is the fastest among all the methods due to small number of basis functions. The total computing time in FRS and DFGPS is smaller than those reported in FRF and DFGPF, since parameter estimation is only done once based on all available training data in the smoothing context; in contrast,  parameter estimation and prediction are done individually for time $u=2, \ldots, 8$, based on data $\bfZ_{1:u}$ in the filtering context. As observations are held out at each time point in the smoothing procedure and only ``current'' observations are held out in the filtering procedure, more observations are held out in the smoothing procedure than those in the filtering procedure. In the implementation of Local Kriging, we only held out observations at each day, and predictions are made based on all remaining observations. DFGPS is faster than Local Kriging. However, more extensive parallelizations can be used to speed up computations in Local Kriging, it will require much more computing resources than FRS and DFGPS.  DFGPS is slow compared to FRS, but it can provide very good inferential results in a reasonable amount of time, since we can obtain predictions at about 316,965 locations from all time points based on about 3.4 million training observations  in a one-week time period. 

\subsection{Filtering and Smoothing Predictions} 
\begin{figure}[htbp]
\begin{subfigure}{.95\textwidth}
  \centering
  \includegraphics[width=1.0\linewidth,height=0.3\textheight]{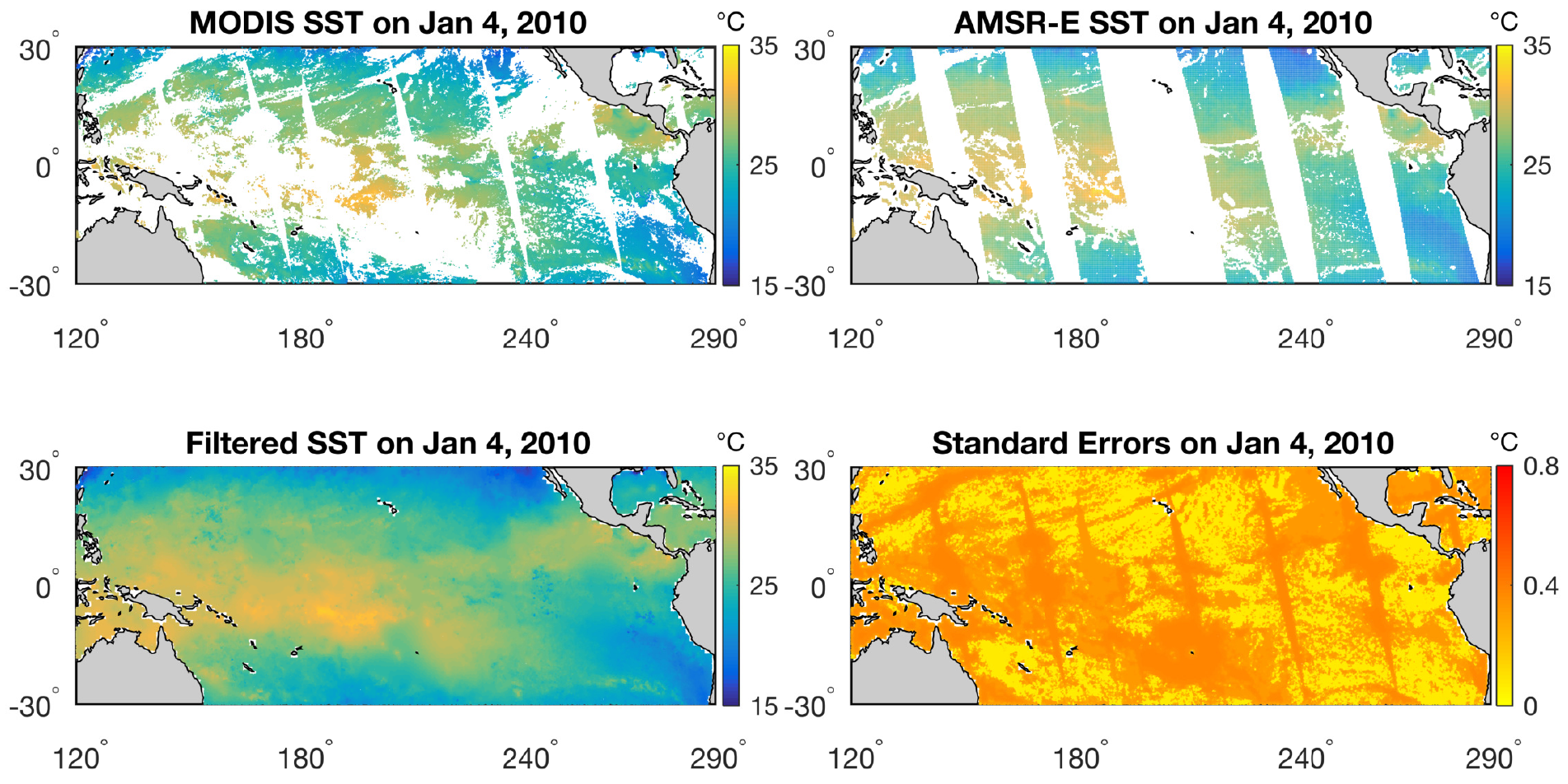}
 % \caption{}
  \label{fig: day4}
\end{subfigure}
\begin{subfigure}{.95\textwidth}
  \centering
  \includegraphics[width=1.0\linewidth,height=0.3\textheight]{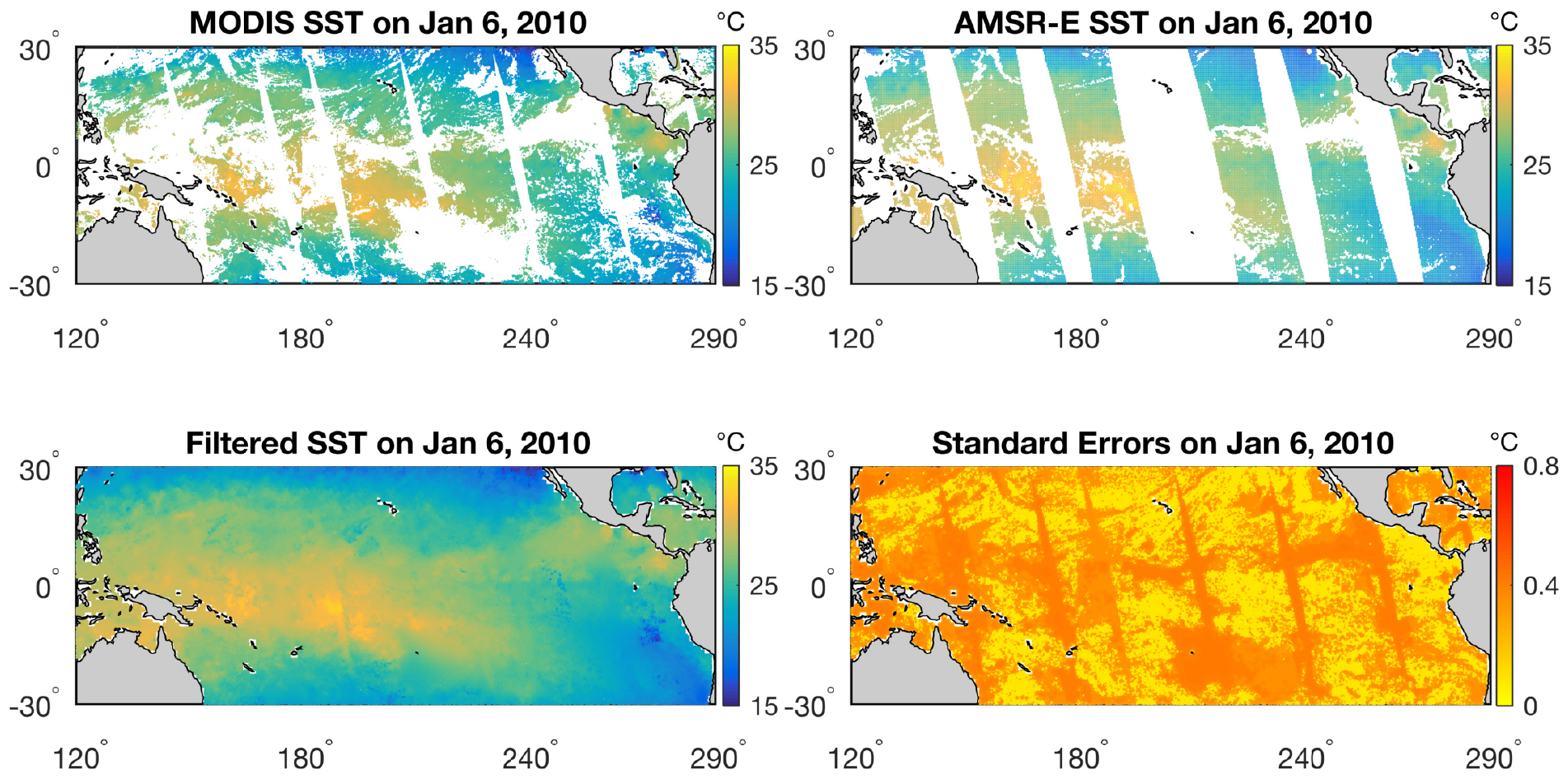}
 % \caption{}
  \label{fig: day6}
\end{subfigure}
\begin{subfigure}{.95\textwidth}
  \centering
  \includegraphics[width=1.0\linewidth,height=0.3\textheight]{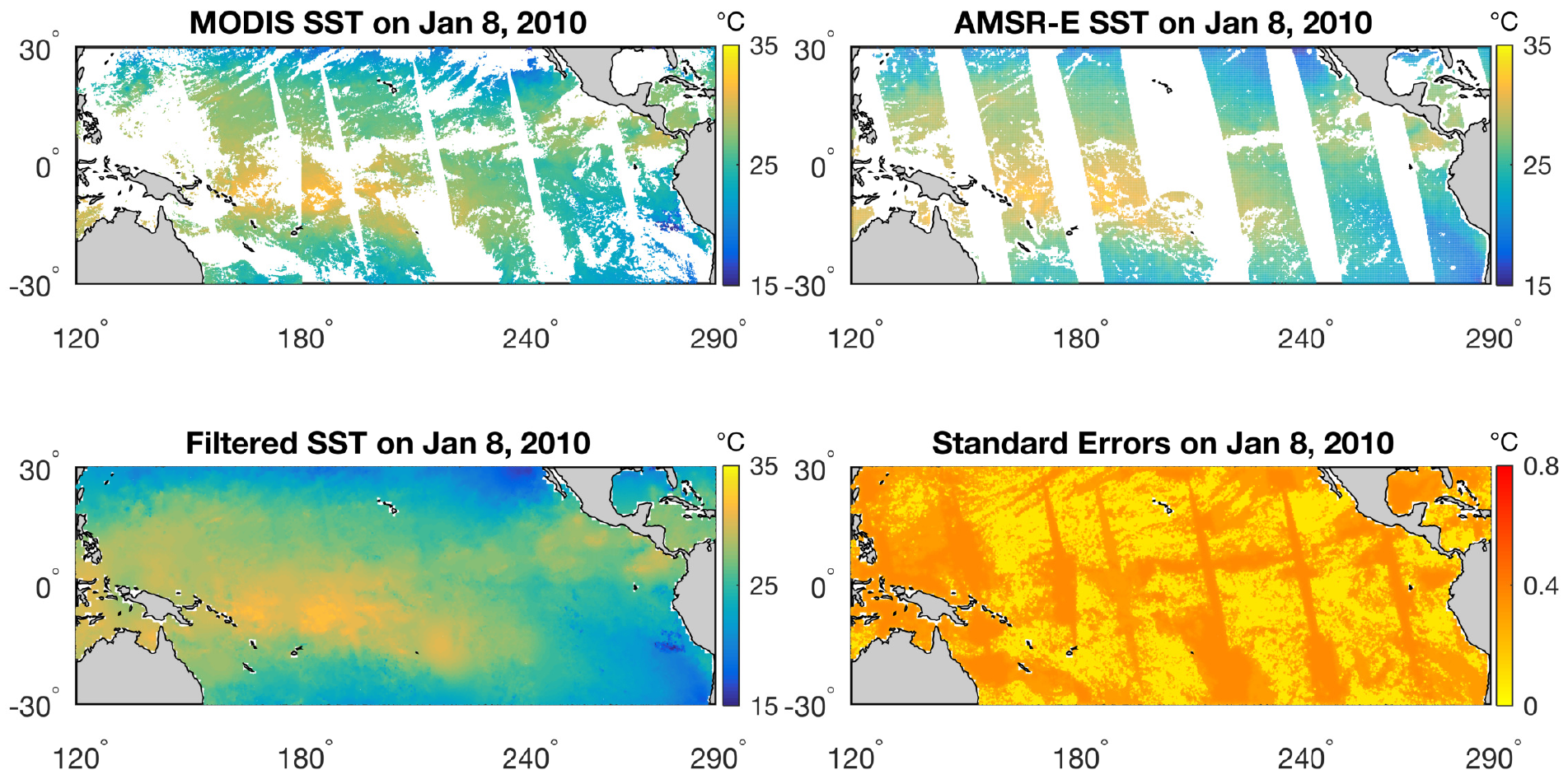}
 % \caption{}
  \label{fig: day8}
\end{subfigure}

\caption{DFGP filtering predictions and associated standard errors on January 4, 6, 8 in the year 2010 over the tropical Pacific ocean.}
\label{fig: predictions in DFGPF}
\end{figure}

\begin{figure}[htbp]
\begin{subfigure}{.95\textwidth}
  \centering
  \includegraphics[width=1.0\linewidth,height=0.3\textheight]{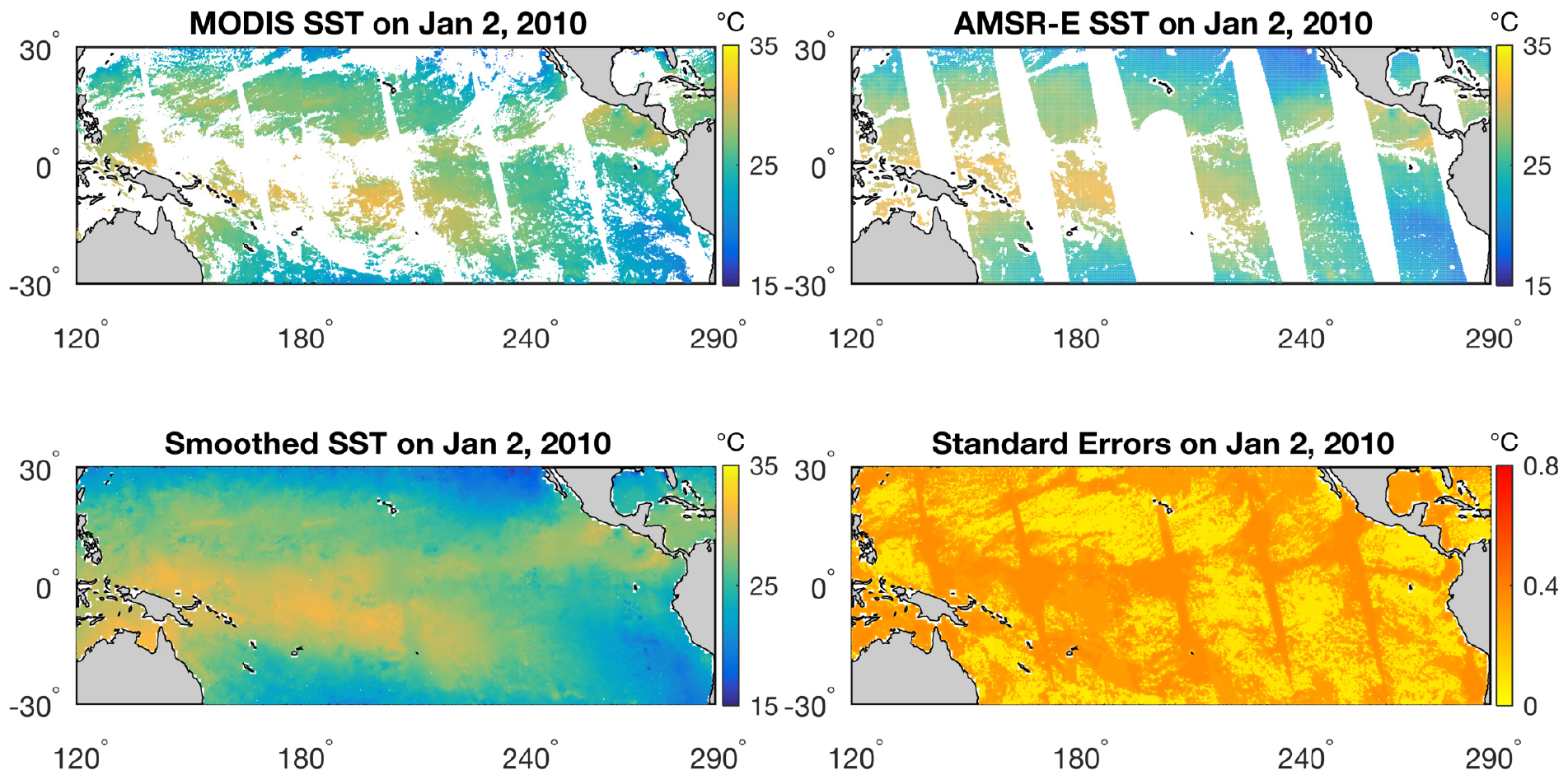}
 % \caption{}
  \label{fig: day4}
\end{subfigure}
\begin{subfigure}{.95\textwidth}
  \centering
  \includegraphics[width=1.0\linewidth,height=0.3\textheight]{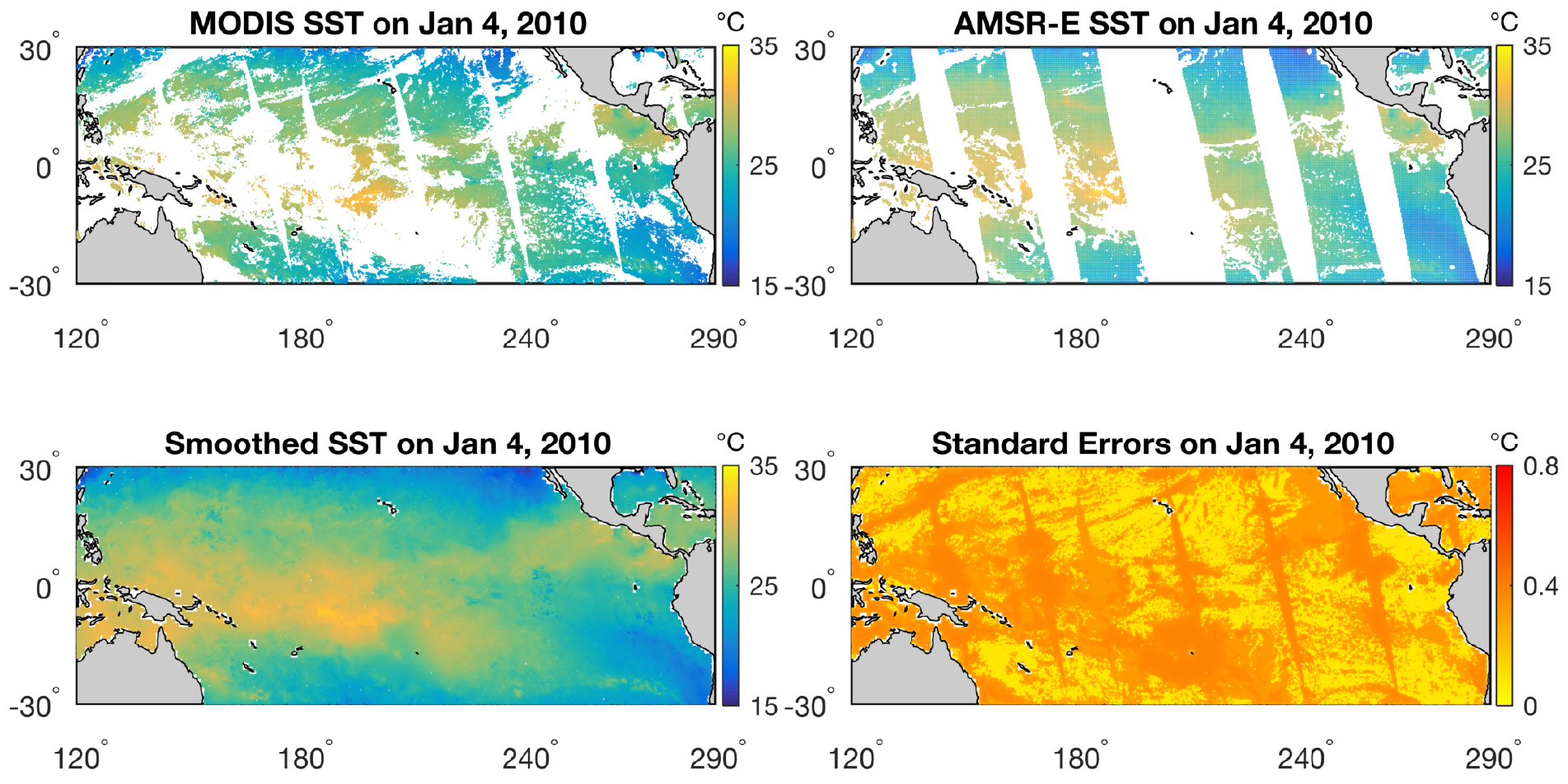}
 % \caption{}
  \label{fig: day6}
\end{subfigure}
\begin{subfigure}{.95\textwidth}
  \centering
  \includegraphics[width=1.0\linewidth,height=0.3\textheight]{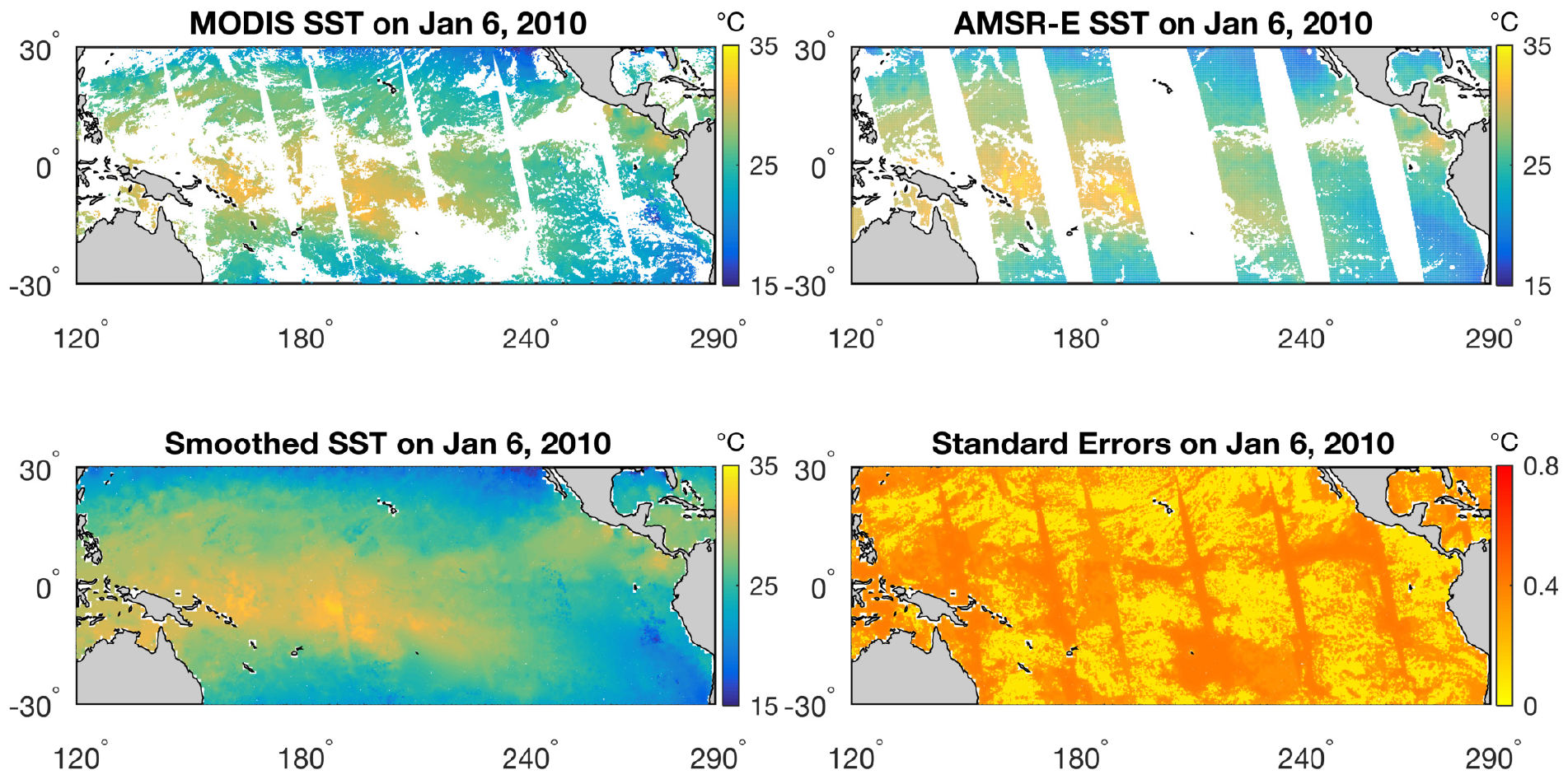}
 % \caption{}
  \label{fig: day8}
\end{subfigure}

\caption{DFGP smoothing predictions and associated standard errors on January 2, 4, 6 in the year 2010 over the tropical Pacific ocean.}
\label{fig: predictions in DFGPS}
\end{figure}

After carrying out cross-validation, we apply DFGPF to make filtering-type predictions for $t=2, \ldots, 8$, and apply DFGPS to make smoothing-type predictions for $t=1, \ldots, 7$. The SEM algorithm with different starting values is used to estimate parameters. It turns out that same parameters were obtained for a pre-specified small threshold after sufficient iterations. This also suggests that the SEM algorithm is robust to initial values as pointed out in \cite{Diebolt1996}. Figure~\ref{fig: predictions in DFGPF} shows the filtering-type predictions for $t= 4, 6, 8$ and associated standard errors, and Figure~\ref{fig: predictions in DFGPS} shows the smoothing-type predictions for $t=2, 4, 6$. As we can see, the resulting predictions are able to fill in the gaps by combining two sources of datasets. The associated prediction standard errors are also reasonable. We see that the predictions show larger uncertainties at locations where no SST data are collected than those at locations where SST data are available. As expected, DFGPS gives better predictions than DFGPF at same time points, since the smoothing methodology makes use of all the available observations. In Supplementary Materials, we also include two movies to show the filtering-type predictions for $t=2, 3, \ldots, 8$, and the smoothing-type predictions for $t=1, 2, \ldots, 7$.

% filtering predictions 

% smoothing prediction

\section{Discussion}\label{sec: conclusion}
In this article, we propose a dynamic fused Gaussian process model to allow both filtering and smoothing type predictions for massive remote sensing data. The parameters in DFGP are estimated via an efficient stochastic expectation-maximization algorithm. The DFGP methodology is demonstrated in a data-fusion context with multiple data sources at different spatial resolutions. We have applied our DFGP model to analyze massive amount of sea surface temperature data from MODIS and AMSR-E satellite instruments in both filtering and smoothing contexts. We found that DFGP gives better prediction results than the spatio-temporal data fusion model in \cite{Nguyen2014} in both filtering and smoothing contexts, even though more basis functions are incorporated. The DFGP methodology also gives much better prediction results than Local Kriging in both filtering and smoothing contexts.

Although DFGP requires more computational cost than Local Kriging, FRF and FRS, the computations in DFGP can be done efficiently with affordable computing resources, since a one-week dataset can be processed in much less than one week for about 3.7 million sea surface temperature observations. By borrowing strength across different time and instruments, DFGP is able to give good prediction results to fill in the gaps for massive amount of sea surface temperature data. A more compelling approach to compare DFGP might be \cite{Jurek2018}. However, it is not clear how this methodology can be extended to a data-fusion context for multiple data sources. The predictions are made at BAU-level, which is motivated by scientific study and available computing resources in practice. A statistical optimal way to choose the resolution of BAUs can be found in \cite{Bradley2017}, but a tradeoff has to be determined between available computing resources and statistical optimality.

The DFGP methodology assumes a single underlying true process, with the data process linked to this true process through different measurement-error processes. The resolution difference among each data process has been explicitly accounted for through the change-of-support property. When different underlying true processes are desired, one can extend the idea in \cite{Nguyen2014} to allow cross dependences among each underlying true process. In the DFGP methodology, the dynamic evolution is only exhibited in the low-rank component, which captures large-scale spatio-temporal variations. Future work might be introducing dynamic evolution structure in the graphical model component.

In this article, the DFGP methodology is demonstrated in a general context without any special structure imposed in the propagation matrix and innovation matrix. In practice, if physical knowledge of a geophysical process is available, this information can be incorporated in the DFGP methodology \cite[e.g.,][]{Wikle2001, Xu2007}. In fact, this will also help avoid potential non-identifiability issue because of over-parameterization in the model when data are sparse. To properly account for uncertainties in both parameter estimation and prediction, a fully Bayesian implementation of the DFGP methodology is recommended, but this will require much more computing resources for massive amount of spatio-temporal data especially for applications in remote sensing science. With high-performance computing facilities, the current DFGP methodology can be applied for much massive amount of data over much larger time period such as the work in \cite{Hoar2003}.

\section*{Acknowledgments}
This work was supported in part by an allocation of computing time from the Ohio Supercomputer Center. 
Ma's research was partially supported by the National Science Foundation under Grant DMS-1638521 to the Statistical and Applied Mathematical Sciences Institute. Any opinions, findings, and conclusions or recommendations expressed in this material do not necessarily reflect the views of the National Science Foundation. {Kang's research was partially supported by the Simons Foundation Collaboration Award (\#317298) and the Taft Research Center at the University of Cincinnati. We thank two anonymous reviewers and an associate editor for comments that greatly improved this work.

\bibliographystyle{apa}

%\singlespacing 
%\setlength{\bibsep}{5pt}

\bibliography{main}{}

\newpage 

\begin{table}[htbp]
\centering
  \caption{Results in the cross-validation study based on the filtering methodology. The RMSPE and CRPS are averaged over all held-out locations and all time points.}
  {\resizebox{1.0\textwidth}{!}{%
  \setlength{\tabcolsep}{3.0em}
  \begin{tabular}{ l c c c c  c  } 
  \toprule \noalign{\vskip 1.5pt} 
		& \multirow{2}*{Local Kriging}  & \multicolumn{2}{c}{FRF} & DFGPF    \\  \noalign{\vskip 1.5pt}  
	 & 	&  $r=99$ & $r=181$ & $r=99$   \\ 
	 \midrule \noalign{\vskip 1.5pt} 
RMSPE  &0.5607	& 0.8354   & 0.7191  & 0.4004 \\ \noalign{\vskip 1.5pt}  \noalign{\vskip 3pt} 
CRPS &0.2587  &0.4402  &0.3751 & 0.1691 \\ \noalign{\vskip 3pt} \hline \noalign{\vskip 3pt}
Time (h) & 20  & 2.4 & 6.5 & 22  \\
\noalign{\vskip .5pt} \bottomrule
  \end{tabular}%
  }}
  \label{table: CV_SST}
\end{table}

\begin{table}[htbp]
\centering
  \caption{Results in the cross-validation study based on the smoothing methodology. The RMSPE and CRPS are averaged over all held-out locations and all time points.}
  {\resizebox{1.0\textwidth}{!}{%
  \setlength{\tabcolsep}{3.0em}
  \begin{tabular}{ l c c c  c  } 
  \toprule \noalign{\vskip 1.5pt} 
		& \multirow{2}*{Local Kriging}  & \multicolumn{2}{c}{FRS} & DFGPS    \\  \noalign{\vskip 1.5pt}  
	 &	&  $r=99$ & $r=181$ & $r=99$   \\ 
	 \midrule \noalign{\vskip 1.5pt} 
RMSPE  & 0.5607 &  0.8270   & 0.7471   & 0.3717  \\ \noalign{\vskip 1.5pt}  \noalign{\vskip 3pt} 
CRPS & 0.2429 & 0.4356  & 0.3873 & 0.1535  \\ \noalign{\vskip 3pt} \hline \noalign{\vskip 3pt}
Time (h) & 20  & 0.9  & 1.4  & 16  \\
\noalign{\vskip .5pt} \bottomrule
  \end{tabular}%
  }}
  \label{table: smoothing CV_SST}
\end{table}

\end{document}